\documentclass[showpacs,twocolumn,nofootinbib]{revtex4}
\usepackage{amsmath}
\usepackage{amsfonts}
\usepackage{amssymb}
\usepackage{graphicx}
\usepackage{epsfig}
\usepackage{epsf}
\begin{document}

\title{A Study of the Evaporative Deposition Process: Pipes and Truncated
Transport Dynamics}
\author{Rui Zheng}
\altaffiliation{Current Address: UBS Global Asset Management, One North Wacker Drive, Chicago, IL 60606}
\email{rui.zheng@ubs.com}
\affiliation{James Franck Institute and Department of Physics, The University of Chicago,
929 E. 57th Street, Chicago, Illinois 60637, USA}
\date{\today}

\begin{abstract}
We consider contact line deposition and pattern formation of a pinned
evaporating thin drop. We identify and focus on the transport dynamics truncated by
the maximal concentration, proposed by Dupont \cite{Dupont1}, as the single deposition
mechanism. The truncated process, formalized as \textquotedblleft pipe
models\textquotedblright, admits a characteristic moving shock front solution
that has a robust functional form and depends only on local conditions. By applying
the models, we solve the deposition process and describe the deposit density profile in different
asymptotic regimes. In particular, near the contact line the density profile
follows a scaling law that is proportional to the square root of the concentration
ratio, and the maximal deposit density/thickness occurs at about $2/3$ of
the total drying time for uniform evaporation and $1/2$ for diffusion-controlled evaporation. 
Away from the contact line, we for the first time identify the power-law decay of the deposit profile with
respect to the radial distance. In comparison, our work is consistent with and extends
previous results \cite{Yuri3}. We also predict features of the
depinning process and multiple-ring patterns within Dupont model, and our
predictions are consistent with empirical evidence. 

\end{abstract}
\maketitle

\section{Introduction}

Evaporative contact line deposition, also known as the \textquotedblleft coffee-drop
effect\textquotedblright, has been the subject of several recent
papers \cite{Deegan1, Deegan4, Deegan2, Deegan3, Yuri4, Yuri3}. The physical
problem originates from a simple phenomenon of everyday life: when a drop
containing a solute such as coffee dries on a surface, the solute is driven to
the contact line and forms a characteristic ring pattern. 
The basic mechanism behind the phenomenon is now well understood \cite{Deegan1}:
the drying drop is pinned at the contact line so that there must be an outward capillary flow to replace
the liquid evaporating at the edge, and this flow carries solute particles to the contact line where the deposit
is formed.     

This simple phenomenon is important in many scientific and
industrial applications \cite{Boneberg1, Joachim1, Rabani1, Magdassi2,
Stone1, Cranick1, Magdassi1}. The evaporation mechanism can create very fine lines of deposit in a
robust way that requires no explicit forming. It can be used to concentrate
material strongly in a controllable way. It also creates capillary flow
patterns which are useful for processing polyatomic solutes like DNA
\cite{Larson1, Larson2, Wong1}.

The deposition process and the resulting patterns are affected by various physical 
and chemical conditions. The wetting property of the substrate and surface roughness 
affect the geometry of an evaporating drop
\cite{Troian1, Doi1}. Also, within the drop there can be a convective flow induced by
gradient in temperature, surfactant concentration, or surface tension
\cite{Danov1, Mitov1, Gennes1, Stebe1}, and it can lead to very complicated, even fractal-like,
deposit patterns. Furthermore, some mechanisms, such as contact line
depinning \cite{Deegan3, Adachi1}, are still not clear, and there are no
satisfactory explanations for multiple-ring deposit patterns \cite{Stone2,
Nonomura1}.

Several models have been established under the common assumptions of slow evaporation
and creeping flow so that quasi-equilibrium dynamics is warranted: the dynamics is determined
by the time evolution of the equilibrium properties of the evaporating system. 
These models use different conditions and assumptions on such 
properties as geometry, evaporation profile, and
deposition criteria. In addition to circular thin drop, there have been studies
on contact line deposition in an angular region \cite{Yuri1, Yuri2, Zheng1},
where the nontrivial geometry at the tip of the angle induces a crossover in
deposit properties. Both uniform evaporation flux and diffusion-controlled
flux, which diverges at the contact line, are considered, 
and they yield different deposit density profiles and growth properties along the contact
line in this asymmetric geometry.

In a circular-symmetric geometry, such as a circular evaporating drop, the
deposit along the contact line is always uniform. It is thus more important
to understand in this case the spatial deposit profile and growth dynamics in a dimension
perpendicular to the contact line, that is, how the solute particles
form the deposit and accumulate toward the center of the drop from the
contact line. Dupont \cite{Dupont1} suggested there be a maximal volume
concentration $C_{\max}$ for solute particles such that once the concentration
increases to $C_{\max}$, the horizontal transport of solute particles stops,
and those particles form the deposit. Following Dupont's
deposition criterion, Popov \cite{Yuri3} studied a model with a pinned round
evaporating thin drop and diffusion-controlled evaporation flux. There are two
regions in his model: a deposition region, where the solute concentration is
$C_{\max}$; a transport region, where the solute concentration is smaller than
$C_{\max}$ and horizontal solute transport is allowed. Popov assumed that the
deposition near the contact line affects the geometry of the drop and the
evaporation-induced flow field (Fig 1.2A). As a result, Popov's model is
characterized by several variables described by a system of coupled
differential equations, of which a complete solution is not available.

\begin{figure}
[ptb]
\begin{center}
\includegraphics[
width=1.0\columnwidth
]%
{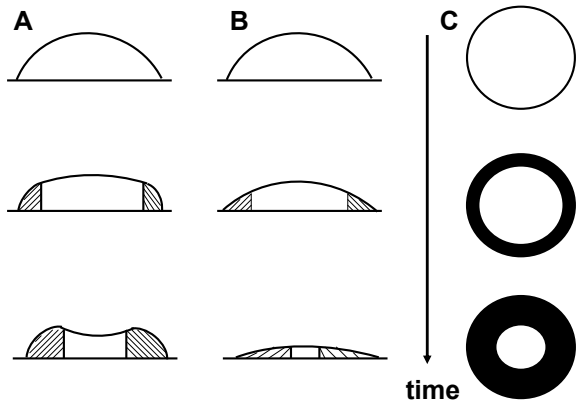}
\caption[Time evolution of a pinned evaporating drop]{Sketches of models of a pinned evaporating drop.
Figure A shows Popov's model, where the deposit (shaded area) maintains its
shape and modifies the geometry of the remaining evaporating drop. Figure B
shows the model considered in this work. The deposit profile does not affect but passively 
changes with the geometry of the evaporating drop. Figure C shows the
projected picture from above. The deposit (dark area) grows from the
contact line toward the center of the drop.}%
\label{two models}%
\end{center}
\end{figure}

In Popov's picture, the deposit thickness increases monotonically
toward the center of the drop until an abrupt vertical wall is formed at the end
of the total drying time. The scale of the thickness is proportional to $\sqrt{\delta_{0}}$, where
$\delta_{0}$ is the concentration ratio defined as the initial uniform volume concentration $C_{0}$
divided by $C_{\max}$. This $\sqrt{\delta_{0}}$-dependence near the contact line had been known before \cite{Deegan3} and
is understandable heuristically. The volume of the deposit, which is proportional to the square of the thickness, 
multiplied by the concentration $C_{max}$, must give the total solute mass in the deposit that is
proportional to $C_{0}$. However, as this dependence is largely asymptotic toward the contact line,
where the time dependence is suppressed and the hydrodynamics and the deposition dynamics are largely separated, 
Popov's coupled system may not be necessary to derive this result. Moreover, the complexity in
Popov's mathematical formulations may inadvertently obscure the simple 
and fundamental underlying physical process. Moreover, a realistic deposit clearly follows a rich profile that shows different
scaling properties in different regions. Thus to interpret the deposit properties within a single asymptotic regime
and with a unique $\sqrt{\delta_{0}}$ is not satisfactory.  

\begin{figure}
[ptb]
\begin{center}
\includegraphics[
width=0.5\columnwidth
]%
{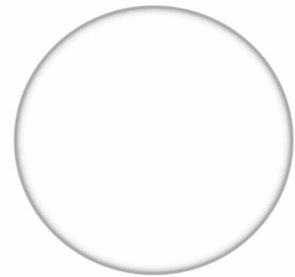}
\caption[Numerical simulation of a ring-like deposit pattern]{Direct numerical simulation of the deposit pattern. The areal
density profile is represented by variation in darkness. (Courtesy Todd
Dupont)}
\label{spot}
\end{center}
\end{figure}

\begin{figure}
[ptb]
\begin{center}
\includegraphics[
width=0.88\columnwidth
]
{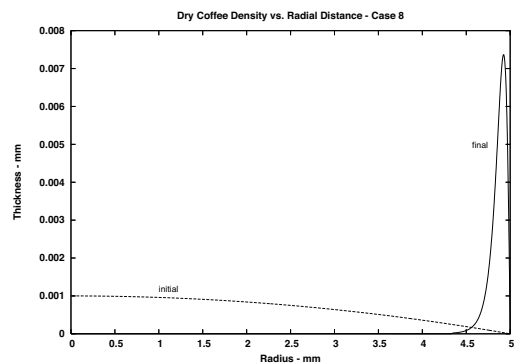}
\caption[Numerical results for the areal density profile]{Areal density profile $\Sigma$ with $C_{0}/C_{\max}=0.002$, where $C_{0}=0.001$ is
the initial uniform volume concentration throughout the drop. The dotted line is the initial areal density identified as
the initial drop height profile multiplied by $C_{0}$. The solid line is the final areal desntity (Courtesy Todd
Dupont)}
\label{profile}%
\end{center}
\end{figure}

Following the truncated transport dynamics and the simpler assumptions
that the hydrodynamics and the deposition process are fully separable, Dupont \cite{Dupont1} 
showed by direct numerical simulation a ring-like deposit pattern
with variance in areal density along the radial direction (Fig. \ref{spot} and
Fig. \ref{profile}). The simple mechanism thus produces a realistic deposit
pattern with a pronounced maximal density near the contact line. In particular, the areal
density shows a potentially analyzable characteristic profile that manifests multiple asymptotic regimes, and that could 
reveal robust properties of contact line deposition under general conditions. 

In this work we re-visit Dupont's criterion with minimal restrictions and
simpler assumptions. We consider a pinned circular thin drop. In
particular, we assume that the deposition process does not interfere with the
hydrodynamics such as the drop geometry and flow velocity field. We consider
both the uniform evaporation and the diffusion-controlled
evaporation conditions, to be defined below. To interpret the deposit density, we assume that
the solute particles in the deposition region, where horizontal transport is
stopped, can still move vertically so that the thickness of the deposit is
determined by the geometry of the drop and changes with time. The final
deposit profile is thus described by its areal density when the drop dries
up and the thickness becomes zero throughout the drop (Fig. 1.2B). To interpret the final
deposit in terms of the areal density profile is not essential to the model,
however, as we shall discuss later that the model is also valid for a deposit
with finite thickness as long as the deposition does not interfere with the
hydrodynamics of the evaporating drop.

We thus want to follow a reductive approach as we believe that the fundamental and robust 
physical properties underlying complex phenomena should be captured by a simple mathematical
structure with minimal restrictions. By this approach, we want to achieve two goals. First, instead of
being exhaustive in considering all the possible conditions, we want to 
understand the robustness of the model: among the
various factors that characterize the problem what are most important in
determining the key deposition properties and in yielding physically feasible and
realistic deposit patterns. Second, what is the appropriate mathematical
structure to describe those dominant factors, and can this mathematical structure 
be generalized to other physical systems.

In the rest of the work we first establish in Section 2 a one-dimensional simplification of 
the deposition process that we denote as \textquotedblleft
pipe models\textquotedblright. These models capture the important characteristics
of the truncated transport dynamics. We then review the major results
of the hydrodynamic properties of an evaporating drop and apply the
pipe models to the evaporative deposition process with uniform evaporation
in Section 3. In Section 4, we solve for the areal density profile and
identify the scaling laws in different asymptotic regimes. We consider in Section 5 the deposition
properties with the diffusion-controlled evaporation. In Section 6, we
compare our findings with Popov's, extending our results to interpreting the deposit
in terms of the thickness profile. We then make a few comments on depinning and formation of multiple-ring
patterns within the truncated transport dynamics scheme in Section 7. We discuss experimental issues
and possible generalizations of the model in Section 8, and present the conclusion in Section 9. In addition,
we provide an alternative derivation of the equation of motion for the shock front in Appendix.

\section{Pipe Models}

The transport process, truncated by the maximal concentration $C_{\max}$, is 
naturally divided into two regions: a transport
region where the concentration is smaller than $C_{\max}$, and a deposition
region where there is no horizontal transport of solute particles. Solute mass is carried
from the transport region to the deposition region, and the growth of the
deposit is described by a characteristic moving front that
separates the two regions.

We establish in this Section the so-called \textquotedblleft pipe
models\textquotedblright\ to formalize the truncated dynamics under different
conditions. Although some predictions made by these models depend on pipe
specifications, we find that certain important properties, such as the shock
front velocity, are actually model-independent and thus generalizable. We
shall show later how these universal properties may help explain the observed
robustness in the general deposition phenomena.

To simplify mathematical formulations, in the following models we only treat
one-dimensional case, or similarly radial symmetric case. However, we believe
the key findings are readily applicable to higher dimensions.

\subsection{A Uniform Pipe}%

\begin{figure}
[ptb]
\begin{center}
\includegraphics[width =0.9\columnwidth]{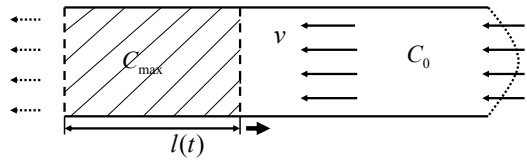}
\caption[Sketch of a uniform pipe]{A uniform pipe. Solute particles with uniform volume
concentration $C_{0}$ is carried at a constant velocity $v$ toward the left
end of the pipe, where deposit is formed with concentration $C_{\max
}>C_{0}$. $l(t)$ is the position of the moving boundary that separates the two regions with different
volume concentration.}
\label{uniform pipeline}
\end{center}
\end{figure}

We consider a uniform pipe with simple properties (Fig.
\ref{uniform pipeline}): incoming solute particles with constant volume
concentration $C_{0}$ are transported at a constant velocity $v$ to the end of
the pipe where the carrier fluid exits the pipe but the solute
particles remain to form a deposit. The solute particles stop moving when their
concentration exceeds $C_{\max}$, which therefore is the volume
concentration of the deposit. To focus on the horizontal transport properties, 
we ignore the variance in velocity along the cross section of the pipe. This simplification is compatible
with the basic assumptions made for the evaporative deposition problem as we shall show in the next Section. 

The configuration is thus divided 
into two regions: a transport region with concentration $C_{0}$ and a deposition region with
concentration $C_{\max}$. The boundary that separates the two regions 
is characterized by its position $l(t) $. Local
conservation of solute mass at the boundary demands:%
\begin{equation}
\frac{dl}{dt}=-v\frac{C_{0}}{C_{\max}-C_{0}}=-v\frac{\delta}{1-\delta},
\label{P1}%
\end{equation}
where $\delta\equiv C_{0}/C_{\max}$.

Several observations immediately follow from Eq. (\ref{P1}). First, the boundary
moves at a velocity proportional to the transport velocity $v$, and the
proportional coefficient is determined by the ratio of the two volumes
concentrations. Second, there must be a finite gap
in volume concentration across the boundary, \textit{i.e.}, $\delta<1$, otherwise
the velocity $dl/dt$ would diverge, which is not physical. The
moving boundary is therefore a shock front that represents a discontinuity in
local volume concentration profile. 

It is worth noting that although the two-component simplified geometry with constant $C_{0}$ and $C_{\max}$ in 
the first place warrants the discontinuity and leads to the shock front, the implications of Eq. (\ref{P1})
are more general. If the transport process is characterized by the truncation criterion and the two-component configuration
and if the boundary that separates the two regions moves with a finite velocity, there gap in concentrations across the
boundary must exist as we now show.

\subsection{A More Realistic Pipe: Non-Singular Case}

\begin{figure}
[ptb]
\begin{center}
\includegraphics[
width=0.90\columnwidth
]%
{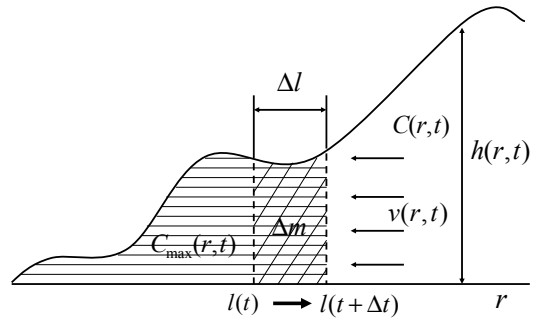}%
\caption[Sketch of a real pipe]{A realistic pipe.
$r$ is the radial distance. The height profile of the pipe is $h(r,t)$. $2\pi rh(r,t)$
characterizes the cross section area, which is not shown, to account for the circularly symmetric
geometry. 
The concentration profile $C(r,t)$ in the transport region, the flow velocity
field $v(r,t)$, and the critical concentration $C_{\max}(r,t)$ are allowed to
change with both the radial distance and time. $l(t)$ is the boundary that separates the
deposition region and the transport region. In a small time interval $\Delta
t$, $l(t)$ moves to $l(t+\Delta t)=l(t)+\Delta l$. $\Delta m$ is the change of
solute mass in the transitional region. }%
\label{real pipeline}%
\end{center}
\end{figure}

A typical evaporating drop that we consider has a circularly
symmetric geometry and an outward flow field along the radial direction. To adapt to this specific
geometry, in a more realistic pipe model (Fig. \ref{real pipeline}) we introduce
$h(r,t)$ as the pipe's height profile, where $r$ is the radial distance, and allow other characteristic
quantities such as the maximal volume concentration $C_{\max}$ and the transport velocity $v$ to vary
both in $r$ and time $t$. In addtion, to account for the circularly symmetric geometry, our pipe
must have a cross section area that varies along the radial direction as $2\pi rh(r,t)$. 

This manipulation with respect to the circularly symmetric geometry
may seem greatly constraint the pipe models. However, we shall show below that the major predictions
made from the pipe models are actually independent of the specific form of the cross section area as long as
it does not introduce significant singularities. The term $2\pi rh(r,t)$ has been introduced
only to conform with the drop geometry which is not really a \textquotedblleft pipe\textquotedblright.

We treat in the real pipe $h$, $C_{\max}$, and $v$ as independent specified physical quantities that have
independent dynamics. This is not necessarily true in real settings such as the evaporative deposition we shall
consider later, but it will allow us to derive general results and will pose not difficulties to implementing these
results to specific real situations where the dynamics may couple. There are perhaps
some conceptual difficulties with the functional form $C_{\max}(r,t)$.
For example, because it evolves in time, what if for a position in the deposition region the $C_{\max}$ 
becomes smaller than the actual volume concentration in a later time? We do not consider this special case as it 
does not occur in the evaporative deposition problem. Nonetheless, it can be relevant in other systems with more complicated
transport properties and is certainly an interesting problem for its own sake.    

We again impose the truncation condition that the horizontal transport of solute mass is truncated once the local volume
concentration $C(r,t)$ reaches $C_{\max}(r,t)$. As in the uniform pipe
model, we consider the position $l(t)$ of the moving front that separates the
deposition region and the transport region. To simplify mathematical analysis,
we assume that all the introduced functions are at lease twice differentiable and that there are no singularities. We
shall however discuss effects of discontinuities and other singularities later.

In the deposition region there is no horizontal transport of solute mass, and
local conservation of mass requires $C(r,t)h(r,t)$ to be time-independent. In
the transport region, solute mass satisfies the equation of continuity
\begin{equation}
\frac{\partial}{\partial t}\left(  Chr\right)  +\frac{\partial}{\partial
r}\left(  vChr\right)  =0. \label{P0}%
\end{equation}

As shown in Fig. \ref{real pipeline}, during the time interval $\Delta t$ the
shock front moves from $l(t)$ to $l(t+\Delta t)=l(t)+\Delta l$. To be consistent 
with the later-introduced context of an evaporating drop, we
choose $\Delta l<0$. $l(t)$ thus decreases in time and moves toward
the right end of the pipe. 

We consider the change of solute mass $\Delta m$ in the
transitional region $\left[  l(t)+\Delta l,l(t)\right] $:

\begin{align}
\Delta m  &  =\Delta m_{1}\nonumber\\
&  =\int_{t}^{t+\Delta t}2\pi l(\tau)\nonumber\\
&  \times\left[  C_{\max}(l\left(  \tau\right)
,\tau)h(l\left(  \tau\right)  ,\tau)-C(l\left(  \tau\right)  ,t)h(l\left(
\tau\right)  ,t)\right] \nonumber\\
&  \times\left(  -\frac{dl}{d\tau}\right)  d\tau, \label{P2}%
\end{align}

\textit{i.e., }$\Delta m$ is equal to the amount of solute mass in the region
when it just becomes a part of the deposition region less the amount of mass in it
when it is still in the transport region. 

We expand $\Delta m_{1}$ up
to the second order of $\Delta t$:

\begin{align}
\Delta m_{1}& =-2\pi l(t)  \nonumber \\
& \times \left[ C_{\max }(l(t),t)h(l(t),t)-C(l(t),t)(h(l(t),t)\right] \frac{%
dl}{dt}\Delta t  \nonumber \\
& -\frac{1}{2}\frac{d}{d\tau }\{2\pi l(\tau )\frac{dl}{d\tau }  \nonumber \\
& \times \left[ C_{\max }(l\left( \tau \right) ,\tau )h(l\left( \tau \right)
,\tau )-C(l\left( \tau \right) ,t)h(l\left( \tau \right) ,t)\right] \left.
\}\right\vert _{\tau =t}\nonumber\\
& \times\left( \Delta t\right) ^{2}+O((\Delta t)^{3})  \nonumber \\
& =-2\pi lh(C_{\max }-C)\frac{dl}{dt}\Delta t  \nonumber \\
& -\pi \left[ \frac{d}{dt}\left( (C_{\max }-C)hl\frac{dl}{dt}\right) +\frac{%
\partial }{\partial t}(Ch)l\frac{dl}{dt}\right] \left( \Delta t\right) ^{2} 
\nonumber \\
& +O((\Delta t)^{3})  \label{P5}
\end{align}

Alternatively, $\Delta m$ is also equal to the amount of solute mass
transported through $l(t+\Delta t)=l(t)+\Delta l$ from time $t$ to $t+\Delta
t$:%
\begin{align}
\Delta m  &  =\Delta m_{2}\nonumber\\
&  =\int_{t}^{t+\Delta t}2\pi l(t+\Delta t)C(l(t+\Delta t),\tau)h\left(
l(t+\Delta t),\tau\right) \nonumber\\
&  \times v\left(  l(t+\Delta t),\tau\right)  d\tau\text{.} \label{P3}%
\end{align}
We also expand $\Delta m_{2}$ up to the second order of $\Delta t$:
\begin{align}
\Delta m_{2} & =2\pi lChv\Delta t+2\pi\left[  \frac{\partial}{\partial
l}(lChv)\frac{dl}{dt}+\frac{1}{2}\frac{\partial}{\partial t}\left(
lChv\right)  \right] \nonumber\\ 
& \times\left(  \Delta t\right)  ^{2}+O((\Delta t)^{3}).\label{P4}
\end{align}

We can now solve for $dl/dt$ given the condition $\Delta m=\Delta m_{1}=\Delta
m_{2}$. If $h(l(t),t)$ is nonvanishing, the terms in the first order of
$\Delta t$ must match, and thus:%

\begin{equation}
-2\pi lh(C_{\max}-C)\frac{dl}{dt}=2\pi lChv, \label{P6}%
\end{equation}%
\begin{equation}
\frac{dl}{dt}=-v\left(  l(t),t\right)  \frac{C(l(t),t)}{C_{\max}%
(l(t),t)-C(l(t),t)}=-v\frac{\delta}{1-\delta}, \label{P7}%
\end{equation}
where $\delta=\delta(r,t)\equiv C(r,t)/C_{\max}(r,t)$.

Eq. (\ref{P7}) has the same form as Eq. (\ref{P1}) in the case of
the uniform pipe though the constant quantities have been replaced here by 
local values of general functions dependent on position and time. The moving front is 
a shock front with $\delta<1$. The front
velocity is proportional to the local transport velocity with opposite
direction, and the proportional coefficient is determined by the local critical
concentration $C_{\max}$ and the local concentration $C$ in the transport
region. Eq. (\ref{P7}) is thus the general non-singular form of the equation of
motion for the shock front in the truncated dynamics.

It also shows that the discontinuity condition $\delta<1$ is not due to the instantiations of constant $C_{0}$
and $C_{\max}$ but the requirement of finite velocity and is inherent in the truncated transport dynamics.
As both concentrations are given functions, Eq. (\ref{P7}) imposes a specific moving boundary $l(t)$ such
that a time-dependent jump in concentration across the boundary is properly maintained and the local solute mass is
thus conserved.

\subsection{A More Realistic Pipe: Singular Case}

If the local height profile $h(l(t),t)$ is zero, the first order term in
$\Delta t$ vanishes for both $\Delta m_{1}$ and $\Delta m_{2}$, and the second order terms in $\Delta t$ thus
must match. The condition $\Delta m=\Delta
m_{1}=\Delta m_{2}$ requires:

\begin{eqnarray}
&&\left[ \frac{d}{dt}\left( (C_{\max }-C)hl\frac{dl}{dt}\right) +\frac{%
\partial }{\partial t}(Ch)l\frac{dl}{dt}\right]   \notag \\
&=&-2\left[ \frac{\partial }{\partial l}(lChv)\frac{dl}{dt}+\frac{1}{2}\frac{%
\partial }{\partial t}\left( lChv\right) \right] ,  \label{P8}
\end{eqnarray}
and
\begin{equation}
\frac{dl}{dt}=-\frac{2\left(  \partial/\partial l\right)  (lChv)\left(
dl/dt\right)  +\left(  \partial/\partial t\right)  (lChv)}{\left(
d/dt\right)  \left(  \left(  C_{\max}-C\right)  hl\right)  +l\left(
\partial/\partial t\right)  \left(  Ch\right)  }. \label{P9}%
\end{equation}

The general form Eq. (\ref{P9}) can be simplified under certain physical
conditions. In particular, we consider the situation where $h(l(t),t)=\partial
h(l(t),t)/\partial t=0$ and $\partial h/\partial l\neq0$, which corresponds to the case 
when the shock front reaches a bottleneck of the pipe. In addition we assume all other
quantities are nonvanishing at the bottleneck. Thus at
$l(t)$%
\begin{equation}
\frac{\partial}{\partial l}(lChv)=Cv\frac{\partial}{\partial l}(hl)\text{,}
\label{P10}%
\end{equation}%
\begin{equation}
\frac{\partial}{\partial t}\left(  Ch\right)  =0, \label{P11}%
\end{equation}
and%
\begin{align}
&\frac{d}{dt}\left(  \left(  C_{\max}-C\right)  hl\right) =\frac{\partial
}{\partial t}\left(  \left(  C_{\max}-C\right)  hl\right)\nonumber\\  
& +\frac{\partial
}{\partial l}\left(  \left(  C_{\max}-C\right)  hl\right)  \frac{dl}%
{dt} =\left(  C_{\max}-C\right)  \frac{dl}{dt}\frac{\partial}{\partial l}\left(
hl\right)  . \label{P12}%
\end{align}
Also, since $C(l(t),t)$ satisfies the equation of continuity (Eq. (\ref{P0})),
\begin{equation}
\left(  \partial/\partial t\right)  (lChv)=-v^{2}C\frac{\partial}{\partial
l}\left(  hl\right)  . \label{P13}%
\end{equation}
Substituting Eqs. (\ref{P10}) to (\ref{P13}), the shock front velocity
Eq. (\ref{P9}) is reduced to
\begin{equation}
\frac{dl}{dt}=-v\frac{C}{C_{\max}-C}\left(  2+\frac{v}{dl/dt}\right)  \text{.}
\label{P14}%
\end{equation}
It can be solved for $dl/dt$:%
\begin{equation}
\frac{dl}{dt}=-v\frac{\sqrt{\delta}}{1-\sqrt{\delta}}, \label{P17}%
\end{equation}
where $\delta=\delta(l(t),t)\equiv C(l(t),t)/C_{\max}(l(t),t)$ is the local
concentration ratio at $l(t)$ such that $h(l(t),t)$ vanishes.

The functional form of Eq. (\ref{P17}) is similar to Eq. (\ref{P7}) for
the nonsingular case with $\delta$ replaced by $\sqrt{\delta}$. However, it is
worth noting that Eq. (\ref{P17}) holds only at those discrete points along the pipe where
the height is zero. In this sense, Eq. (\ref{P17}) is not a differential equation but a boundary
condition or a series of boundary conditions determined by the properties of the pipe in question.
We shall come back to this point in Discussion.

\subsection{Truncated Transport Dynamics and the Shock Front}

We have shown that the truncated dynamics in various pipe models is
described by a moving shock front $l(t)$, of which the equation of motion is
determined by the local values of the characteristic quantities at the front
under the condition of mass conservation. We now establish the formal
mathematical framework for one-dimensional truncated transport dynamics
and derive the equation of motion for the characteristic shock front directly.
We only treat non-singular case and omit rigorous discussion of boundary and
initial conditions. In particular, we do not consider the problem \emph{how the
dynamics is initiated} but once it has been established \emph{what the equation of motion
should be}. The solution to the former depends on the boundary and initial
conditions but is less relevant to applying the models to the
evaporative deposition problem.

We consider a one-dimensional transport equation which imposes both the
continuity condition with respect to the generalized mass $\varphi(r,t)C(r,t)$
and the truncation condition imposed on $C(r,t)$:%

\begin{align}
&\frac{\partial}{\partial t}(\varphi(r,t)C(r,t))\nonumber\\
& +\frac{\partial}{\partial
r}(v(r,t)\varphi(r,t)C(r,t)\theta(C_{\max}(r,t)-C(r,t))=0, \label{A1}%
\end{align}
where $v(r,t)$ is the velocity field, $C_{\max}(r,t)$ is the truncation threshold, and
$\theta$ is the step function defined by:%

\begin{equation}
\theta(x)=%
\genfrac{\{}{.}{0pt}{}{1\text{, }x>0,}{0\text{, }x\leq0.}
\label{A8}%
\end{equation}

For general purposes, $\varphi\left(  r,t\right)  $, $C\left(  r,t\right)  $,
and $C_{\max}\left(  r,t\right)  $ in Eq. (\ref{A1}) have no specific
physical meaning attached though $\varphi(r,t)$ can be regarded as the ratio
between the generalized mass and the quantity on which the truncation
condition is imposed. In the case of the pipe models we have considered,
$C(r,t)$ is the solute volume concentration and $\varphi\left(  r,t\right)
=rh(r,t)$ where $h\left(  r,t\right)  $ is the height profile of the pipe. For
non-singular solutions, we assume $\varphi(r,t)$, $v(r,t)$, and $C_{\max
}(r,t)$ are positive and at least twice differentiable.

\begin{figure}
[ptb]
\begin{center}
\includegraphics[
width = 0.90\columnwidth
]%
{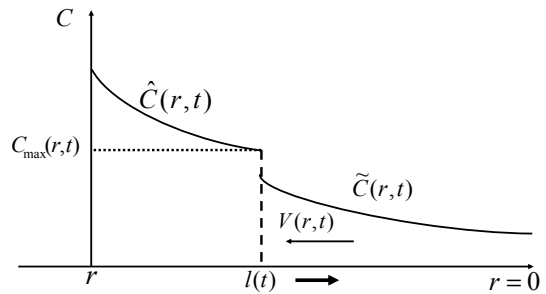}%
\caption[Sketch of the truncated dynamics solution]{Sketch of the truncated dynamics solution at time $t$. The solution
consists two parts: $\tilde{C}(r,t)$ in the transport region and $\hat
{C}(r,t)$ in the deposition region. The two regions are separated by the shock
front $l(t)$. There is a nonzero jump in $C(r,t)$ across the shock front. }%
\label{truncated math}%
\end{center}
\end{figure}

We seek solution that admits a moving shock front $l(t)$:%
\begin{equation}
C(r,t)=\tilde{C}(r,t)+\theta(l(t)-r)(\hat{C}(r,t)-\tilde{C}(r,t))\text{,}
\label{A3}%
\end{equation}
where $\tilde{C}(r,t)$ and $\hat{C}(r,t)$ are again at least twice
differentiable, $\tilde{C}(r,t)<C_{\max}(r,t)$ for $r\geq l(t)$, and $\hat
{C}(l(t),t)=C_{\max}$. We show the sketch of $C(r,t)$ in Fig.
\ref{truncated math}.

Substituting Eq. (\ref{A3}) into Eq. (\ref{A1}), all singular terms must
cancel each other. Thus,  
\begin{equation}
(\varphi(\hat{C}-\tilde{C})\frac{dl}{dt}+v\varphi\tilde{C})\delta(l-r)=0,
\label{A4}%
\end{equation}
and, given $\varphi(l(t),t)$ is not zero,
\begin{equation}
\frac{dl}{dt}=-v(r,t)\frac{\tilde{C}(l(t),t)}{C_{\max}(l(t),t)-\tilde
{C}(l(t),t)}. \label{A5}%
\end{equation}
Eq. (\ref{A5}) is the general result of pipe models, and in the non-singular case it
does not depend on $\varphi(r,t)$.

The coffee stain problem we shall discuss in the next Section corresponds to the specific case
where $\varphi(r,t)=rh(r,t)$, $C_{\max}(r,t)=constant$, and $C(r,0)=C_{0}.$

\section{Evaporative Deposition with Uniform Evaporation}

The evaporative deposition process is a special case of the pipe
models above: hydrodynamics of the thin evaporating drop determines a height profile
$h(r,t)$ and generates a flow velocity field $v(r,t)$; solute particles with
initial uniform volume concentration $C_{0}$ are carried by the
evaporation-generated flow toward the contact line, and the transport process
is truncated by a maximal allowed concentration $C_{\max}$; once the volume
concentration reaches $C_{\max}$, the horizontal transport is stopped and
deposit forms.

The volume concentration profile of solute particles is thus divided into two regions:

\emph{Transport region}. Solute particles move with the evaporation-induced
flow at the same velocity and are transported toward the contact line along the radial direction. The
volume concentration in this region solves the continuity equation with a finite
velocity field.

\emph{Deposition region.} The horizontal transport of solute particles is
truncated, and only vertical movement is allowed afterward. The solute mass
at each position, $i.e.$, $C(r,t)h(r,t)$ is thus conserved in time. All the solute
particles eventually deposit at the position and form the areal density profile at
the total drying time. We also assume that the deposition process is
decoupled from the hydrodynamics of the evaporating drop, and both the height
profile $h(r,t)$ and the velocity field $v(r,t$) are not affected by the deposition process.

There must be a moving shock front $l(t)$ that separates these
two regions. The velocity of the front $dl/dt$ must obey the predictions made
by the pipe models.

To apply the pipe models, we need to specify the height profile $h(r,t)$, the velocity field $v(r,t)$, 
and the solute volume concentration in the transport region. Most of these have already been explored in
the existent literature \cite{Deegan3, Yuri3} though some exact results are not immediately available. For the sake of
completeness, we shall nonetheless go through the whole derivation process from our own perspective.

\subsection{Review of Evaporating Thin Drop Hydrodynamics}

\begin{figure}
[ptb]
\begin{center}
\includegraphics[
width = 0.95\columnwidth
]
{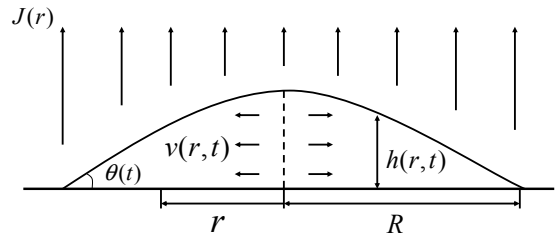}
\caption[Configuration of an evaporating thin drop]{Sketch of an evaporating thin drop. We assume the drop is pinned during 
the time under consideration. The radius of the drop $R$ is thus a constant. The geometry 
is described by a height profile $h(r,t)$ that depends on the radial distance and time. $\theta(t)$ 
is the contact angle that decreases linearly in
time in our approximation. $J(r)$ is the evaporation flux profile. $v(r,t)$ is
the flow velocity field that also depends on the radial distance and time.}
\label{thin drop}
\end{center}
\end{figure}

We consider a circular evaporating thin liquid drop (Fig. \ref{thin drop}). The
drop is pinned in the drying process, and the drop radius $R$ is thus a
constant. The geometry of the drop is determined by the contact angle
$\theta(t)$ and the height profile $h(r,t)$.

Under the assumption of slow evaporation, the quasi-equilibrium condition can be
applied so that at any time the drop maintains mechanical equilibrium against the pressure
difference across the liquid-air interface.\footnote{This condition holds
as long as the flow velocity within the drop is much smaller than the characteristic velocity $\sigma/3\eta$ where
$\sigma$ is the surface tension and $\eta$ is the dynamic viscosity \cite{Yuri3}. For water the characteristic
velocity is about 24 m/s.} The shape of the drop is thus determined by
\begin{equation}
\Delta p=-2H\sigma\text{,} \label{T1}%
\end{equation}
where $\Delta p=\Delta p(t)$ is the pressure difference, $H=H(t)$ is the mean
curvature of the drop surface, and $\sigma$ is the surface tension. It is well
known that for a circular drop the geometry that solves Eq. (\ref{T1}) is a
spherical cap with the contact angle $\theta$ determined by:%

\begin{equation}
\sin\theta(t)=\frac{R}{2}\frac{\Delta p(t)}{\sigma}. \label{T2}%
\end{equation}
The height profile $h(r,t)$ in this case is%
\begin{align}
h(r,t)  &  =\sqrt{\frac{R^{2}}{\sin^{2}\theta(t)}-r^{2}}-R\cot\theta
(t)\nonumber\\
&  \approx\frac{R^{2}-r^{2}}{2R}\theta(t),\text{ \ }\theta\ll1\text{.}
\label{T3}%
\end{align}
If the evaporation flux does not change with time, it is confirmed
experimentally \cite{Deegan3} that when $\theta$ is small,\footnote{The initial contact angle $\theta_{i}$ in
Deegan's experiments was about 0.1 to 0.3 radians \cite{Deegan3}.} the contact angle $\theta(t)$ decreases 
linearly with time for the most part of the drying process:
\begin{equation}
\theta(t)=\theta_{i}(1-\frac{t}{t_{f}}), \label{T4}%
\end{equation}
where $\theta_{i}$ is the initial contact angle and $t_{f}$ is the total
drying time that depends on specific forms of evaporation profile. In what follows
we shall restrict our discussion to thin drops so that the linear relation Eq. (\ref{T4})
is warranted.

Loss of solvent due to evaporation generates inside the drop a flow field
$\mathbf{u}(r,z,t)$ with horizontal component $u_{s}(r,z,t)$, where $z$
represents the coordinate along the vertical direction. We introduce a
vertically averaged velocity field
\begin{equation}
v(r,t)=\frac{1}{h(r,t)}\int_{0}^{h(r,t)}u_{s}(r,z,t)dz. \label{T5}%
\end{equation}
Local conservation of solvent mass then demands%
\begin{equation}
\nabla\cdot(hv)+\frac{J(r)}{\rho}\sqrt{1+\left(  \nabla h\right)  ^{2}%
}+\partial_{t}h=0, \label{T6}%
\end{equation}
where $\rho$ is the density of the solvent, and $J(r)$ is the evaporation profile
defined as mass loss per unit projected area per unit time. For a thin circular
drop $v(r,t)$ is along the radial direction and $\left\vert \nabla h\right\vert
\ll1$, and Eq. (\ref{T6}) can be integrated to solve for $v$:%
\begin{equation}
v(r,t)=-\frac{1}{rh}\int_{0}^{r}\left(  \frac{J}{\rho}+\partial_{t}h\right)
rdr\text{.} \label{T7}%
\end{equation}
For uniform evaporation in which $J(r)=const.$, it can be shown that:%
\begin{equation}
v(r,t)=\frac{1}{4}\frac{r}{t_{f}}\left(  1-\frac{t}{t_{f}}\right)  ^{-1}.
\label{T8}%
\end{equation}
where the total drying time $t_{f}$, when the whole drop dries up via
evaporation, is 
\begin{equation}
t_{f}=\frac{\rho}{\pi R^{2}J}\int_{0}^{R}2\pi h(r,0)rdr=\frac{\theta_{i}R\rho
}{4J}. \label{T90}%
\end{equation}

To simplify mathematical formulations, we choose $R$ as the unit of length,
$t_{f}$ as the unit of time, and $R/t_{f}$ as the unit of velocity. Thus in this model
the dynamics of an evaporating drop is characterized by two dimensionless functions:
\begin{equation}
h(r,t)=\frac{\theta_{i}}{2}(1-r^{2})(1-t), \label{T10}%
\end{equation}%
\begin{equation}
v(r,t)=\frac{1}{4}r(1-t)^{-1}, \label{T11}%
\end{equation}
where we have used the same notations $r$ and $t$ to represent the dimensionless distance and time
respectively for simplicity.

In deriving $v(r,t)$ (Eq. (\ref{T11})), we have used the condition that the mass of the solvent is conserved locally
(Eq. (\ref{T6})), and this condition involves $h(r,t)$. Thus $v(r,t)$ and $h(r,t)$ have coupled
dynamics as opposed to the general situation we considered in the real pipe models. This coupling will
not pose any difficulty, however.

\subsection{Solute Volume Concentration Profile}

\subsubsection{Concentration Profile in the Transport Region}

\begin{figure}
[ptb]
\begin{center}
\includegraphics[
width = 0.9\columnwidth
]%
{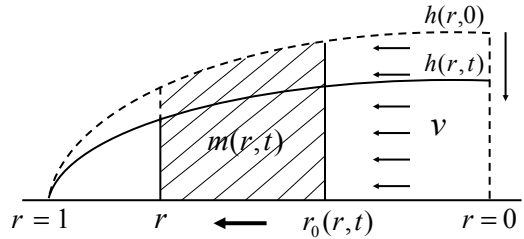}%
\caption[Sketch of the definition of $m(r,t)$]{Sketch of the definition of $m(r,t)$. A\ solute particle at $r_{0}$
at time zero moves to $r$ at time $t$. $r_{0}$ as a function of $r$ and $t$ is
given by Eq. (\ref{S2}). $m(r,t)$ is equal to the total solute mass between
$r_{0}$ and $r$ (the shaded area) at time zero.}%
\label{transport concentration}%
\end{center}
\end{figure}

We now derive the volume 
concentration $C(r,t)$ in the transport region, assuming the defined height profile
(Eq. (\ref{T10})) and velocity field (Eq. (\ref{T11})). 

For radial distance $r$ and time $t$ we define $r_{0}(r,t)$ to be the initial position such that
solute particles start at time zero from $r_{0}$ reach $r$ at time $t$.
$r_{0}$ can be solved via integrating the velocity field (Eq. (\ref{T11})):
\begin{equation}
r_{0}(r,t)=r(1-t)^{1/4}. \label{S2}%
\end{equation}

We next define $m(r,t)$ to be the amount of solute passing through $r$ from time zero to
time $t$ (Fig. \ref{transport concentration}). Then $m(r,t)$ is equal to the total amount of solute between $r$ and
$r_{0}(r,t)$ at time zero:
\begin{align}
m(r,t) & =2\pi\int_{r_{0}(r,t)}^{r}C_{0}h(r',0)r'dr'\nonumber\\
& =\pi C_{0}\theta_{i}\left[
\frac{r^{2}}{2}\left(  1-\left(  1-t\right)  ^{1/2}\right)  -\frac{r^{4}}%
{4}t\right]  . \label{S3}%
\end{align}

\begin{figure}
[ptb]
\begin{center}
\includegraphics[
width = 0.9\columnwidth
]
{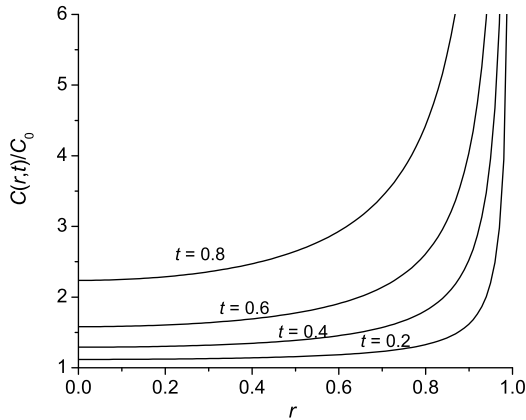}%
\caption[Volume concentration profile in the transport region]{The solute volume concentration profile $C(r,t)/C_{0}$ (Eq.
(\ref{S7})) in the transport region at $t=0.2$, $0.4$, $0.6$, and $0.8$. }%
\label{density profile}%
\end{center}
\end{figure}

We consider a ring-like region between $r$ and $r+\Delta r$. At time zero the
total amount of solute mass in this region is
\begin{equation}
m_{0}=2\pi r\Delta rh(r,0)C_{0}. \label{S4}%
\end{equation}
At time $t$, it becomes
\begin{equation}
m_{t}=2\pi r\Delta rh(r,t)C(r,t). \label{S5}%
\end{equation}
The change of solute mass up to time $t$ in this region should
be equal to the total amount of mass moving into $r$ less the total amount of
mass moving out of $r+\Delta r$, 
\begin{equation}
\Delta m=m(r,t)-m(r+\Delta r,t). \label{S6}%
\end{equation}
Conservation of solute mass demands $m_{0}+\Delta m=m_{t}$, thus%
\begin{align}
C(r,t)  &  \equiv\tilde{C}(r,t) \nonumber\\
& =\lim_{\Delta r\rightarrow0}\left(  C_{0}%
\frac{h(r,0)}{h(r,t)}+\frac{m(r,t)-m(r+\Delta r,t)}{2\pi r\Delta
rh(r,t)}\right) \nonumber\\
&  =C_{0}\frac{\left(  1-t\right)  ^{-1/2}-r^{2}}{1-r^{2}}, \label{S7}%
\end{align}
where, in accordance with the notations in pipe models, we use $\tilde
{C}(r,t)$ to denote the volume concentration in the transport region. As shown
in Fig. \ref{density profile}, $\tilde{C}(r,t)
$ increases monotonically with both $r$ and $t$.%

The concentration profile (Eq. (\ref{S7})) is a \textquotedblleft leaking
solution\textquotedblright\ in the sense that the total amount of solute mass in
the transport region is not conserved, and rightly so. This can be
demonstrated in an ideal evaporative deposition process such that all the
solute particles are eventually carried to the contact line without truncation
in volume concentration \textit{i.e.}, in the limit $C_{\max}\rightarrow
\infty$. It is verifiable that $\tilde{C}(r,t)$ satisfies the continuity
equation (Eq. (\ref{P0})):%
\begin{equation}
\frac{\partial}{\partial t}\left(  \tilde{C}hr\right)  +\frac{\partial
}{\partial r}\left(  v\tilde{C}hr\right)  =0. \label{S20}%
\end{equation}
The total amount of solute mass at time $t$ in the body of the drop is
\begin{equation}
M(t)=\int_{0}^{1}2\pi\tilde{C}hrdr. \label{S21}%
\end{equation}
Thus the rate of change of the total solute mass is,
\begin{align}
\frac{dM(t)}{dt} & =\int_{0}^{1}2\pi\frac{\partial}{\partial t}(\tilde
{C}hr)dr=-2\pi\int_{0}^{1}\frac{\partial}{\partial r}\left(  v\tilde
{C}hr\right)  dr\nonumber\\
& =\left.  2\pi v\tilde{C}hr\right\vert _{r=1}^{r=0},
\label{S23}%
\end{align}
that is, the change of the total mass is equal to the amount of mass injected into
the drop (at $r=0$) less the amount of mass flowing out of the drop boundary
(at $r=1$) that immediately forms deposit at the contact line. The transport
process thus must be truncated with some finite threshold $C_{\max}$ so that a moving
boundary is formed and  the deposit grows from the contact line toward 
the center of the drop.

\subsubsection{Concentration Profile in the Deposition Region}

We are now ready to apply the pipe models. There is a shock front $l(t)$
that separates the transport region and the deposition region. The equation of
motion for the shock front is given by Eq. (\ref{P7}) for the nonsingular
case, where $\delta=C(r,t)/C_{\max}$ is the concentration ratio at the front.
Substituting the height profile $h(r,t)$ (Eq. (\ref{T10})), the flow velocity
field (Eq. (\ref{T11})), and the volume concentration profile in the transport
region (Eq. (\ref{S7})), the velocity $dl/dt$ is given by:%

\begin{equation}
\frac{dl}{dt}=-v\frac{\delta}{1-\delta}=-\frac{l}{4(1-t)}\frac{1}%
{\frac{C_{\max}\left(  1-l^{2}\right)  }{C_{0}\left[  (1-t)^{-1/2}%
-l^{2}\right]  }-1}. \label{P16}%
\end{equation}

To determine $l(t)$ completely, in addition to Eq. (\ref{P16}), we also need to
specify initial conditions. At time zero, the shock front $l(t)$ starts at the
contact line: $l(0)=1$. Also, physically $dl/dt$ must have a
well-define value at time zero, but naively for $t=0$ and $l(0)=1$ the right
side of Eq. (\ref{P16}) is not well defined. The pipe models again can
shed some light on this. At contact line the height is zero, and the pipe models 
in the singular case (Eq. (\ref{P17})) require
\begin{equation}
\left.  \frac{dl}{dt}\right\vert _{t=0}=-v\left(  0,0\right)  \frac
{\sqrt{\delta_{0}}}{1-\sqrt{\delta_{0}}}=-\frac{1}{4}\frac{\sqrt{\delta_{0}}%
}{1-\sqrt{\delta_{0}}},\text{ }l(0)=1, \label{P18}%
\end{equation}
where $\delta_{0}=C_{0}/C_{\max}$. Eqs. (\ref{P16}) and (\ref{P18}) thus
completely define the time evolution of the shock front $l(t)$.

The seeming discontinuous \textquotedblleft jump\textquotedblright in $dl/dt$ from 
the $\sqrt{\delta_{0}}$-dependence at the contact line to $\delta$-dependence once $l$ moves
away is artificial. The concentration profile in the transport region $\tilde{C}(r,t)$
(Eq. (\ref{S7})) has no definite limit at the contact line for $r = 1$ and $t = 0$.
Specific limit thus must be calculated along a definite path $r = l(t)$. The transition from
$\sqrt{\delta_{0}}$ to $\delta$ is actually continuous along this path. This can be
demonstrated directly (see Appendix for details), and will be further clarified later when we
discuss the deposit profile properties near the contact line.

Once $l(t)$ is known, the concentration profile in the deposition region can
be determined under the condition that $C(r,t)h(r,t)$ is conserved in this region. Thus for $r>l(t)$
\begin{equation}
C(r,t)\equiv\hat{C}(r,t)=C_{\max}\frac{h(r,l^{-1}(r))}{h(r,t)}, \label{S14}%
\end{equation}
where $l^{-1}$ is the inverse function of $l$: $l^{-1}(r)$ gives the time when
the shock front arrives at the position $r$. $^{\text{ }}$

Combining the concentration profile $\tilde{C}$ (Eq. (\ref{S7})) in the
transport region and the concentration profile $\hat{C}$ (Eq. (\ref{S14})) in
the deposition region via the moving the shock front $l(t)$, the total
concentration profile is
\begin{equation}
C(r,t)=\tilde{C}(r,t)+\theta(r-l(t))(\hat{C}(r,t)-\tilde{C}(r,t))\text{.}
\label{S30}%
\end{equation}

Alternatively the shock front velocity (Eq. (\ref{P16})) and the boundary
condition (Eq. (\ref{P18})) can be derived directly via solute mass
conservation. We show the derivation in details in Appendix.

\subsection{The Shock Front}

The region boundary $l(t)$ is a moving shock front, and there is a finite
gap in volume concentration across the boundary.

Naively we can introduce a virtual moving front $\tilde{l}(t)$ by
extrapolating the concentration profile (Eq. (\ref{S7})) in the transport
region with the condition $\tilde{C}(\tilde{l}(t),t)=C_{\max}$. Thus
\begin{equation}
\tilde{l}(t)=\left[  \frac{1}{1-\delta_{0}}-\frac{\delta_{0}}{1-\delta_{0}%
}\left(  1-t\right)  ^{-1/2}\right]  ^{1/2}\text{,} \label{S15}%
\end{equation}
where $\delta_{0}=C_{0}/C_{\max}$. $\tilde{l}(t)$ represents the location of
the moving front if there were no discontinuity in volume concentration. 
Evidently $t\leq1-\delta_{0}^{2}$ for Eq. (\ref{S15}) to be meaningful, that is, the virtual
front $\tilde{l}(t)$ arrives at the center of the drop before the whole drop
dries up at $t=1$. 

We can also derive the equation of motion
for $\tilde{l}(t)$:%
\begin{equation}
\frac{d\tilde{l}}{dt}=-v(\tilde{l}(t),t)\frac{\delta_{0}}{\left(  1-t\right)
^{1/2}-\delta_{0}}, \label{S17}%
\end{equation}%
\begin{equation}
\left.  \frac{d\tilde{l}}{dt}\right\vert _{t=0}=-\frac{1}{4}\frac{\delta_{0}%
}{1-\delta_{0}}. \label{S18}%
\end{equation}
Thus the virtual front moves slower than the shock front starting from the contact line.

Since for $0<t<1$ the volume concentration $C(r,t)$ (Eq. (\ref{S7})) in the transport region increases with $r$
and at the shock front $C(l(t),t)<C_{\max}=C(\tilde{l}(t),t)$, thus
\begin{equation}
l(t)<\tilde{l}(t), \label{S16}%
\end{equation}
for $0<t<1-\delta_{0}^{2}$.

\begin{figure}
[ptb]
\begin{center}
\includegraphics[
width=0.9\columnwidth
]
{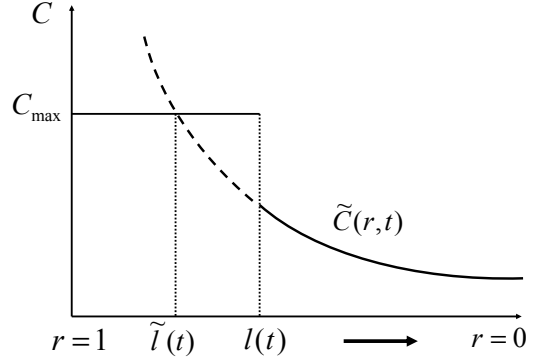}
\caption[Shock front versus virtual front]{Sketches of the shock front and the virtual front at time $t$. The
shock front is ahead of the virtual front: $l(t)<\tilde{l}(t)$. There is a
nonzero jump in volume concentration across the shock front: $C_{\max}-C(l(t),t)$,
with $C_{\max}=C(\tilde{l}(t),t)$. }
\label{virtual front}
\end{center}
\end{figure}

\begin{figure}
[ptb]
\begin{center}
\includegraphics[
width=1.0\columnwidth
]%
{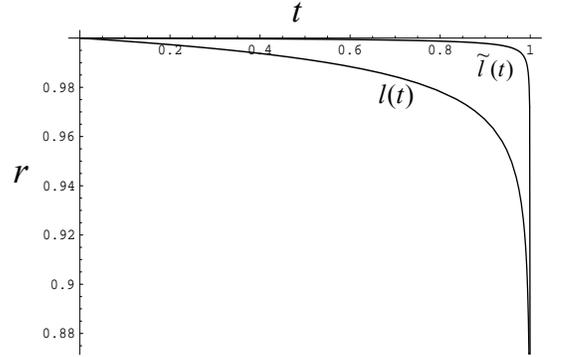}%
\caption[Time evolutions of the two fronts]{Time evolution of the two fronts with $\delta_{0}=0.002$. $l(t)$ and
$\tilde{l}(t)$ start from the contact line ($r=1$) at time zero and arrive at the
center of the drop ($r=0$) at time $t=1-\delta_{0}^{2}$.}%
\label{two fronts}%
\end{center}
\end{figure}

We know that $l(0)=\tilde{l}(0)=1$, and we shall prove later that
$l(1-\delta_{0}^{2})=\tilde{l}(1-\delta_{0}^{2})=0$. The shock front and the
virtual front thus start out from the contact line at time $t=0$ with the shock
front moving faster. The virtual front then catches up,
and both fronts arrive at the center of the drop at time $t=1-\delta_{0}^{2}.$
The sketches of these two fronts at time $t$ are shown in Fig.
\ref{virtual front}, and their time evolutions are shown in Fig.
\ref{two fronts}. We note that both fronts spend most of the total drying time near the contact line ($l\rightarrow1
$).

\section{Deposit Density Profile with Uniform Evaporation}

\subsection{Areal Density}

When a drop dries up, a ring-like stain is left with nonzero width and
variation in density. The spatial profile of the deposit pattern is
characterized by the areal density $\Sigma$, defined as the amount of solute mass per
unit area. For a pinned circular drop without irregularities, $\Sigma$ is a
function of the radial distance $r$.

Since in the deposition region there is no horizontal solute transport, at any
position in this region all the solute particles distributing vertically
above the position along the height of the drop will eventually deposit and form the areal
density at the position. We consider a position with radial distance $r$ from the center of the drop. At time
$l^{-1}(r)$ when the shock front arrives at $r$, the drop height at $r$ is $h(r,l^{-1}(r))$, 
and the volume concentration is exactly $C_{\max}$. Thus substituting the expression for the
height profile (Eq. (\ref{T10})), we find
\begin{equation}
\Sigma(r)=C_{\max}h(r,l^{-1}(r))=\frac{\theta_{i}C_{\max}}{2}(1-r^{2}%
)(1-l^{-1}(r)). \label{d1}%
\end{equation}
Alternatively, via the shock front function $l(t)$ the density profile $\Sigma$
can also be expressed as a function of time
\begin{equation}
\Sigma(t)=\frac{\theta_{i}C_{\max}}{2}(1-l^{2}(t))(1-t), \label{d20}%
\end{equation}
representing the growth dynamics of the deposit. We shall use the same functional 
notation $\Sigma$ for both the areal density
profile $\Sigma(r)$ and the growth dynamics $\Sigma(t)$ interchangeably without confusion.

\begin{figure}
[ptb]
\begin{center}
\includegraphics[
width=0.91\columnwidth
]%
{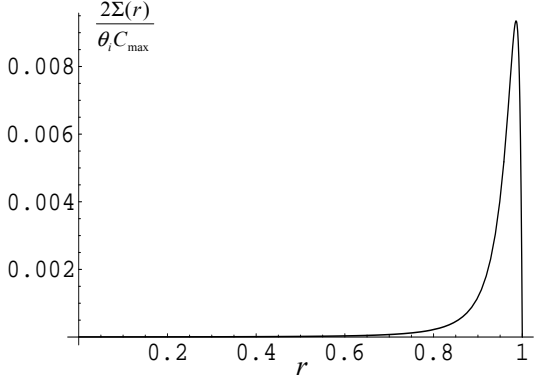}%
\caption[Areal density profile]{The areal density profile $2\Sigma(r)/(\theta_{i}C_{\max})$ from the
center of the drop ($r=0 $) to the contact line ($r=1$) with $\delta
_{0}=0.002$, obtained by numerical solution of Eqs. (\ref{P16}) and (\ref{d1}).}%
\label{sigma_r}%
\end{center}
\end{figure}

\begin{figure}
[ptb]
\begin{center}
\includegraphics[
width=0.88\columnwidth
]%
{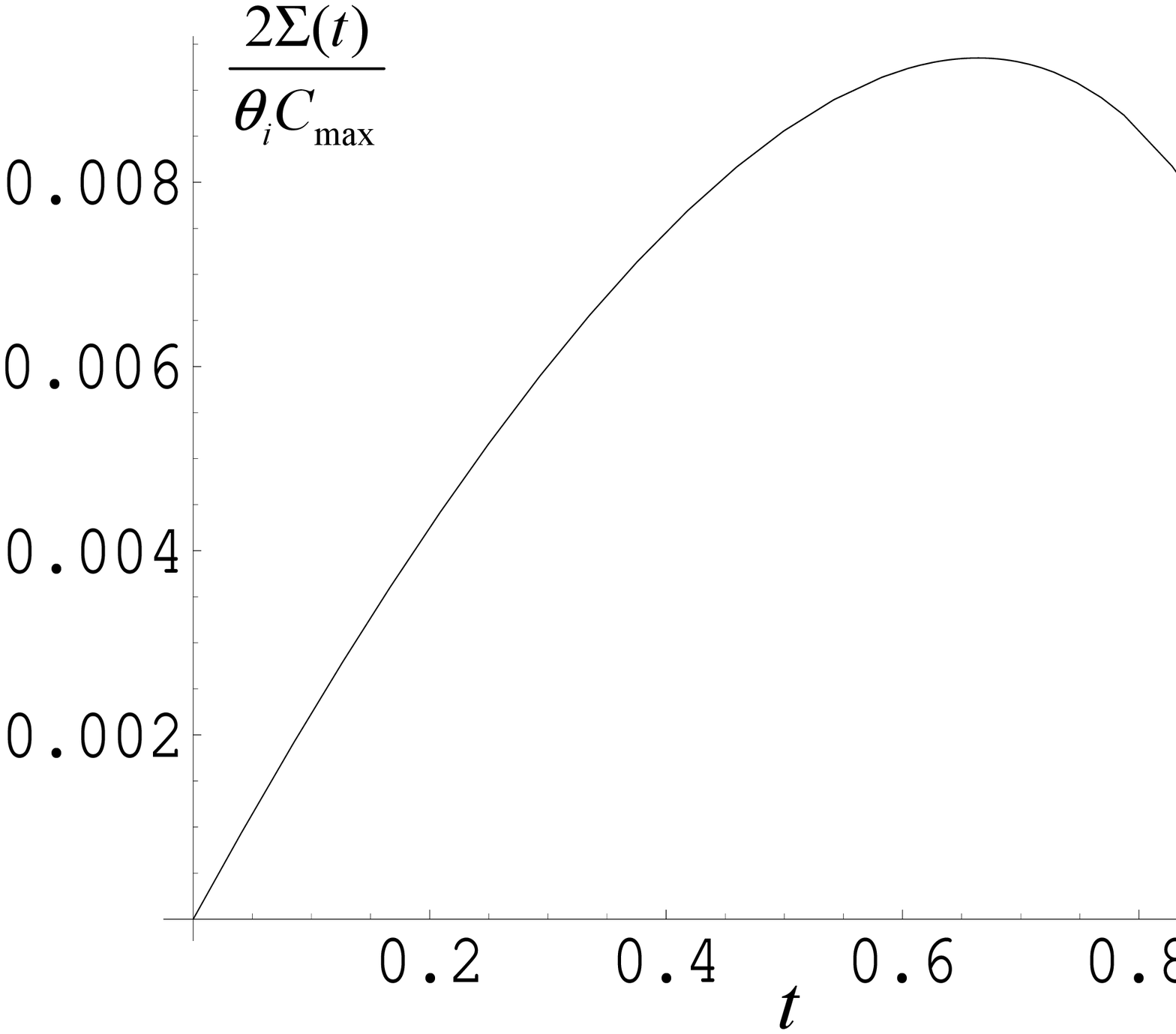}%
\caption[Areal density growth dynamics]{Areal density growth dynamics $2\Sigma(t)/(\theta_{i}C_{\max})$ with
$\delta_{0}=0.002$. The time domain is essentially from $0$ to $1-\delta
_{0}^{2}$.}%
\label{sigma_t}%
\end{center}
\end{figure}

A direct numerical solution for $l(t)$ (Eq. (\ref{P16})) yields an areal
density profile $\Sigma\left(  r\right)  $ shown in Fig. \ref{sigma_r} and an
areal density growth $\Sigma\left(  t\right)  $ shown in Fig. \ref{sigma_t}.
The areal density profile shows that almost all the deposit occurs near the
contact line with a pronounced peak followed by a steep
decrease and then a long flat tail with negligible deposit. The areal
density growth dynamics $\Sigma\left(  t\right)  $, however, shows that
a significant amount of deposit occurs throughout the whole drying process
with the maximal density much less pronounced in the time profile than the
spatial profile. This is understandable from previous results (Fig.
\ref{two fronts}) as the shock front $l(t)$ remains near the contact line for
the most of the total drying time.%

We now study these profiles analytically via solving the solving for the shock front $l(t)$. 
In the small $\delta_{0}$ limit we seek the solution to Eq. (\ref{P16}) via
asymptotic expansion. Since $0\leq t<1$, we expand $l(t)$ in terms of $t$ and thus find
\begin{equation}
l(t)=1-\frac{1}{4}\sqrt{\delta_{0}}t-\frac{7}{48}\sqrt{\delta_{0}}t^{2}%
-\frac{61}{576}\sqrt{\delta_{0}}t^{3}+O(t^{4}). \label{d4}%
\end{equation}

We look for the position $r_{m}$ and the time $t_{m}=l^{-1}(r_{m})$ for the
maximal areal density near the contact line. Numerical simulations show%
\begin{equation}
1-r_{m}^{N}=0.31...\sqrt{\delta_{0}},\ t_{m}^{N}=0.67.... \label{d7}%
\end{equation}
Using first order approximation in $t$: $l(t)\approx1-(1/4)\sqrt
{\delta_{0}}t$, we find
\begin{equation}
1-r_{m}^{(1)}=0.125...\sqrt{\delta_{0}},\text{ \ }t_{m}^{(1)}=0.5. \label{d5}%
\end{equation}
Second and third order approximations in $t$ yield:%
\begin{equation}
1-r_{m}^{(2)}=0.18...\sqrt{\delta_{0}},\text{ }t_{m}^{(2)}=0.55..., \label{d8}%
\end{equation}%
\begin{equation}
1-r_{m}^{(3)}=0.22...\sqrt{\delta_{0}},\text{ }t_{m}^{(3)}=0.59.... \label{d9}%
\end{equation}

It seems that different orders of approximation all show a contact line distance
$1-r_{m}$ proportional to $\sqrt{\delta_{0}}$ but a maximal deposition time $t_{m}$ independent of $\delta_{0}$,
and that with higher order of approximation the resulting $1-r_{m}$ and $t_{m}$ converge to the numerical results
Eq. (\ref{d7}). We thus conjecture that, in the small $\delta_{0}$ limit, the distance between
the maximal areal density and the contact line scales with $\sqrt{\delta_{0}}%
$, and the time at which the maximal areal density occurs is a constant
portion of the total drying time. We shall confirm this conjecture later.

We also note that since throughout the drying process the majority of the
deposit occurs near the contact line, we can analyze the deposit spatial
profile $\Sigma\left(  r\right)  $ and the growth dynamics $\Sigma\left(  t\right)  $
separately in different asymptotic regimes.

\subsection{Asymptotic Time-Independent Deposition}

We consider two cases where the time evolution of the deposition
characteristics is negligible either by focusing on the depositing process in
the early time regime or by controlling the experimental conditions to assure time independence.

\subsubsection{Deposition at Initial Drying Stage}

The initial drying stage corresponds to the asymptotic regime $t/t_{f}\rightarrow0$ 
or equivalently $t_{f}\rightarrow\infty$. Similar to the previous study \cite{Zheng1}, 
we consider a typical
evaporative deposition scenario where there is a large pool of solution
($t_{f}\rightarrow\infty$), and we are only interested in the deposition
process near the contact line ($r\rightarrow1$) up to a time $t/t_{f}$ close to zero.

In this approximation the time dependence of all the characteristic quantities
is suppressed. Thus with $t\rightarrow0$ and $r\rightarrow1$
\begin{equation}
h(r,t)\approx\frac{\theta_{i}}{2}(1-r^{2})\approx\theta_{i}\left(  1-r\right)
, \label{F1}%
\end{equation}%
\begin{equation}
v(r,t)\approx v(1,0)\equiv v_{0}, \label{F2}%
\end{equation}%
\begin{equation}
\delta\approx\delta_{0}=\frac{C_{0}}{C_{\max}}, \label{F3}%
\end{equation}%
and
\begin{equation}
l(t)\approx1-\sqrt{\delta_{0}}v_{0}t. \label{F100}
\end{equation}
The deposit profile satisfies:%
\begin{equation}
\Sigma(r)\approx\theta_{i}C_{\max}\left(  1-r\right)  , \label{F4}%
\end{equation}%
\begin{equation}
\Sigma(t)\approx\theta_{i}C_{\max}\left(  1-l(t)\right)  \approx\theta
_{i}C_{\max}\sqrt{\delta_{0}}v_{0}t. \label{F5}%
\end{equation}

The spatial profile (Eq. (\ref{F4})) is trivial as it is just the time-independent 
drop height profile (Eq. (\ref{F1})) multiplied by the
maximal concentration $C_{\max}$. The growth of deposit from the contact
line is linear in time with the proportional coefficient determined by the
initial conditions of the drying drop.

Eq. (\ref{F4}) also shows, at the initial drying stage,
\begin{equation}
\frac{d\Sigma(r)}{dr}=-\theta_{i}C_{\max}. \label{F25}%
\end{equation}
The slope of the deposit profile near the contact line only depends on the
initial contact angle and the critical concentration $C_{\max}$. We note that
in this case the initial flow velocity $v_{0}$ is finite at
the contact line, and this is not a general property. We shall later discuss
the case of the diffusion-controlled evaporation profile where $v_{0}$
diverges, and we shall show the slope (Eq. (\ref{F25})) is actually independent of the
evaporation profile. Intuitively, this property is not difficult to understand. Since the time-dependence is suppressed, the slope of the areal density at
the contact line should be identical/proportional to the slope of the initial drop height
profile when the drying process begins, which is exactly the initial contact angle $\theta_{i}$.

\subsubsection{Deposition in Time-Independent Geometry}

\begin{figure}
[ptb]
\begin{center}
\includegraphics[
width=0.93\columnwidth
]%
{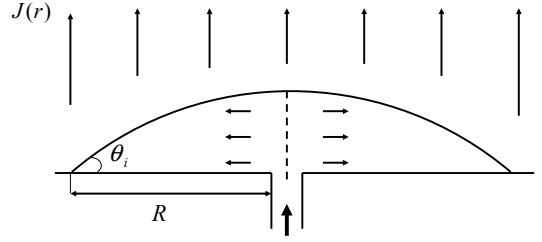}%
\caption[Sketch of a time-independent geometry]{Sketch of a possible realization of time-independent geometry.
Solvent is injected into the drop at the center from below to balance the loss
due to evaporation.}%
\label{constant geometry}%
\end{center}
\end{figure}

An evaporating drop with time-independent geometry can be achieved by
replacing the evaporating fluid as illustrated in Fig. (\ref{constant geometry}). 
Solvent is injected into the drop from below at the
drop center with solvent mass $\int_{0}^{R}2\pi rJ(r)dr$ per unit time so that
the total amount of solvent in the drop is constant in time. The manipulation
also allows minimal perturbations to the drop shape and the flow velocity field
and thus achieves a drop height profile that is approximately time-independent.%

\begin{figure}
[ptb]
\begin{center}
\includegraphics[
width=0.88\columnwidth
]%
{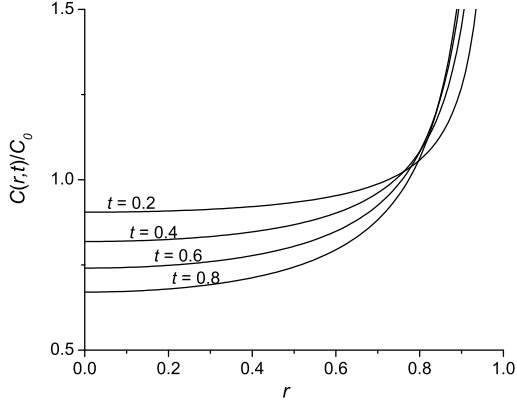}%
\caption[Concentration profile in a time-independent geometry]{Volume concentration profile $C(r,t)/C_{0}$ (Eq. (\ref{F9})) in the
time-independent geometry at $t=0.2$, $0.4$, $0.6$, and $0.8$, with $\delta_{0}=0.002$.}%
\label{density profile_constant time}%
\end{center}
\end{figure}

The hydrodynamics in this case is thus stationary, and with uniform
evaporation profile
\begin{equation}
h(r)=h(r,t=0)=\frac{\theta_{i}}{2}\left(  1-r^{2}\right)  , \label{F6}%
\end{equation}%
\begin{equation}
v(r)=v(r,t=0)=\frac{1}{4}r. \label{F8}%
\end{equation}
Using the same method (Eqs. (\ref{S2}), (\ref{S3}), and (\ref{S7})), we can show that the volume concentration profile in
the transport region is:%
\begin{equation}
\tilde{C}(r,t)=C_{0}\frac{\exp\left(  -t/2\right)  -r^{2}\exp\left(  -t\right)
}{1-r^{2}}, \label{F9}%
\end{equation}
where $C_{0}$ is the initial uniform volume concentration. We show the volume
concentration profiles $C(r,t)/C_{0}$ at four different times in Fig.
\ref{density profile_constant time}. As opposed to the time-dependent geometry,
the concentration increases with time near the contact line but 
decreases close to the drop center, and there is a crossover.

The spatial areal density profile is trivial:
\begin{equation}
\Sigma(r)=C_{\max}h(r)=\frac{\theta_{i}C_{\max}}{2}(1-r^{2}),
\label{F10}%
\end{equation}
and the deposit growth dynamics is
\begin{equation}
\Sigma\left(  t\right)  =\frac{\theta_{i}C_{\max}}{2}\left(  1-l^{2}%
(t)\right)  , \label{F11}%
\end{equation}
where the evolution of the shock front $l(t)$ is given by the pipe models Eq. (\ref{P7}).

\subsection{Density Profile in Spatial Asymptotic Regimes}

\subsubsection{Near the Contact Line}

We now analyze the areal density profile in spatial asymptotic regimes. We
first consider the limit $l(t)\rightarrow1$. In addition we require,
\begin{equation}
\delta_{0}(1-t)^{-1/2}\ll1,
\label{d30}%
\end{equation}
that is, we consider in the time regime not too late in the whole drying
process as we shall show later that the shock front reaches the drop center
exactly at $1-\delta_{0}^{2}$. Evidently this is satisfied for any $t<1$ for
sufficiently small $\delta_{0}$.

According to the previous results for the deposit profile $\Sigma\left(
r\right)  $ and the growth process $\Sigma\left(  t\right)  $ (Fig. \ref{sigma_r}
and Fig. \ref{sigma_t}), almost all the deposit occurs throughout almost
all the drying process in this asymptotic regime. Thus this regime contains most
characteristics of the deposition profile. In particular, we want to
look at where (the contact line distance $(1-r_{m})$) and when (time $t_{m}$) the
maximal areal density occurs, and to show explicitly the scaling properties with
respect to $\sqrt{\delta_{0}}$.

In this asymptotic regime Eq. (\ref{P16}) is reduced to
\begin{equation}
\frac{dl}{dt}=-\frac{\delta_{0}}{4(1-t)}\frac{\left(  1-t\right)  ^{-1/2}%
-1}{1-l^{2}}. \label{d10}%
\end{equation}
With the initial condition Eq. (\ref{P18}), $l(t)$ can be solved for:
\begin{equation}
\left(  1-l\left(  t\right)  \right)  ^{2}=\frac{\delta_{0}}{4}\left[
2\left(  1-t\right)  ^{-1/2}+\ln\left(  1-t\right)  -2\right]  . \label{d11}%
\end{equation}
We substitute Eq. (\ref{d11}) into $\Sigma(t)$ (Eq. (\ref{d20})) 
and look for the maximal value. We find the time $t_{m}$ satisfies 
the following equation

\begin{equation}
2\ln(1-t_{m})+3\left(  1-t_{m}\right) ^{-1/2}-3=0. \label{d12}%
\end{equation}

Thus
\begin{equation}
t_{m}\approx0.67, \label{d13}%
\end{equation}%
\begin{equation}
r_{m}=l(t_{m})=1-0.31...\sqrt{\delta_{0}}, \label{d14}%
\end{equation}%
and
\begin{equation}
\Sigma_{m}=\Sigma\left(  r_{m}\right)  =0.11...\sqrt{\delta_{0}}\theta_{i}C_{\max}. \label{d15}%
\end{equation}

The maximal areal density occurs roughly at $2/3$ of the total drying time and
is independent of the density ratio $\delta_{0}$ when it is small. The
distance from the contact line to the maximal areal density $1-l_{m}$ and the
maximal density itself, on the other hand, scale with $\sqrt{\delta_{0}} $.
These results are consistent with our preliminary findings via asymptotic expansion.

\begin{figure}
[ptb]
\begin{center}
\includegraphics[
width=0.86\columnwidth
]%
{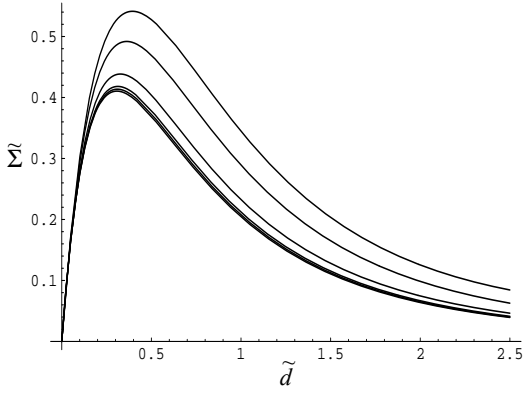}
\caption[Scaling property of the areal density profile for small $\delta_{0}$]{Scaling property of the areal density profile for small $\delta_{0}$.
The scaled areal density is defined as $\tilde{\Sigma}=2\Sigma/(\theta
_{i}R\sqrt{\delta_{0}})$; the scaled radial distance is defined as $\tilde
{d}=\left(  1-r\right)  /\sqrt{\delta_{0}}$. The spectrum of curves going downward corresponding to respectively
$\delta_{0}=0.1$, $0.05$, $0.01$, $0.002$, $0.001$, and $0.0005$.}
\label{scaling law_r}%
\end{center}
\end{figure}

We can show the scaling properties more explicitly. Near the contact line
$r\rightarrow1$ and $1-r^{2}=(1+r)(1-r)\approx2(1-r)$ (Eq. (\ref{d1})), and thus
$\Sigma(r)\approx\theta_{i}C_{\max}(1-r)(1-l^{-1}(r))$. We introduce the scaled
quantities
\begin{equation}
\tilde{\Sigma}=\frac{\Sigma}{\theta_{i}C_{\max}\sqrt{\delta_{0}}} \label{d21}%
\end{equation}
and 
\begin{equation}
\tilde{d}=1-\tilde{r}=\frac{1-r}{\sqrt{\delta_{0}}}, \label{d22}%
\end{equation}
and Eq. (\ref{d20}) shows
\begin{equation}
1-t=\frac{\tilde{\Sigma}}{\tilde{d}}. \label{d23}%
\end{equation}
Substituting Eq. (\ref{d23}) into Eq. (\ref{d11}), we find
\begin{equation}
\tilde{d}^{2}=\frac{1}{4}\left[  2\left(\frac{\tilde
{d}}{\tilde{\Sigma}}\right)^{1/2}+\ln\frac{\tilde{\Sigma}}{\tilde{d}%
}-2\right]  . \label{d24}%
\end{equation}
Eq. (\ref{d24}) implicitly gives the scaled universal areal density profile
$\tilde{\Sigma}$ as a function of the scaled contact line distance
$\tilde{d}$ (Fig. \ref{scaling law_r}). 

Eq. (\ref{d24}) cannot be solved explicitly, and we instead look for the
asymptotic power laws in different limits. We first consider the limit
$\tilde{d}\rightarrow0$. On the right side of Eq. (\ref{d24}), the term
containing logarithm dominates, and thus
\begin{equation}
\tilde{\Sigma}\sim\tilde{d}e^{4\tilde{d}^{2}}\approx\tilde{d}.\label{d26}%
\end{equation}
The areal density increases with the contact line distance linearly. Eq.
(\ref{d26}) is consistent with our previous findings (Eq. (\ref{F4})).

We next consider the tail of the profile in the limit $\tilde{d}%
\rightarrow\infty$ and $\tilde{\Sigma}\rightarrow0$. In this limit
$(\tilde{d}/\tilde{\Sigma})^{1/2}$ is the dominant term, and thus%
\begin{equation}
\tilde{\Sigma}\sim\frac{1}{4}\tilde{d}^{-3}.\label{d27}%
\end{equation}
In terms of the original varibles $\Sigma$ and $r$, Eq. (\ref{d27}) can be
rewritten as%
\begin{equation}
\Sigma\sim\frac{1}{4}\theta_{i}C_{\max}\frac{\delta_{0}^{2}}{\left(
1-r\right)  ^{3}}.\label{d28}%
\end{equation}
The areal density profile is thus proportional to $\delta_{0}^{2}$ along the
tail as opposed to $\sqrt{\delta_{0}}$ toward the contact line.   

\begin{figure}
[ptbptb]
\begin{center}
\includegraphics[
width=0.88\columnwidth
]%
{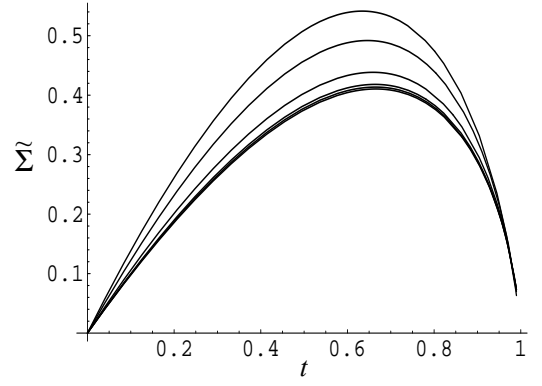}
\caption[Scaling property of the areal density growth dynamics for small
$\delta_{0}$]{Scaling property of the areal density growth dynamics for small
$\delta_{0}$. The scaled areal density is defined as $\tilde{\Sigma}%
=2\Sigma/(\theta_{i}R\sqrt{\delta_{0}})$. The spectrum of curves going downward corresponding to respectively
$\delta_{0}=0.1$, $0.05$, $0.01$, $0.002$, $0.001$, and $0.0005$.}%
\label{scaling law_t}%
\end{center}
\end{figure}

Similarly we can show the scaled areal density $\tilde{\Sigma}$ as a function
of time $t$ explicitly:%
\begin{equation}
\tilde{\Sigma}=\frac{1-t}{2}\left[  2\left(  1-t\right)  ^{-1/2}+\ln\left(
1-t\right)  -2\right]  ^{1/2}. \label{d25}%
\end{equation}
Eq. (\ref{d25}) gives the universal dynamics of deposit growth (Fig. \ref{scaling law_t}).

\subsubsection{Toward the Center of the Drop}

Toward the center of the drop $l\rightarrow0$, Eq. (\ref{P16}) takes the form%
\begin{equation}
\frac{dl}{dt}=-\frac{l}{4(1-t)}\frac{\delta_{0}\left(  1-t\right)  ^{-1/2}%
}{1-\delta_{0}\left(  1-t\right)  ^{-1/2}}\text{.} \label{d16}%
\end{equation}
We find in this case
\begin{equation}
l=A\left[  1-\delta_{0}\left(  1-t\right)  ^{-1/2}\right]  ^{1/2}, \label{d17}%
\end{equation}
where $A$ is a constant and can not be determined without proper boundary conditions.

Thus toward the center of the drop $l(t)\propto\tilde{l}(t)$ (Eq.
(\ref{S15})), and as a result
\begin{equation}
l(1-\delta_{0}^{2})=\tilde{l}\left(  1-\delta_{0}^{2}\right)  =0\text{.}
\label{d18}%
\end{equation}
The shock front arrives at the center of the drop at time $(1-\delta_{0}%
^{2})t_{f}$, and the whole horizontal transport process stops before the drop dries up.

The areal density in this case is%
\begin{equation}
\Sigma(r)\propto\delta_{0}^{2}\frac{A^{4}(1-r^{2})}{\left(  A^{2}%
-r^{2}\right)  ^{2}}\approx\delta_{0}^{2}\left[  1+\left(  \frac{2}{A^{2}%
}-1\right)  r^{2}\right]  . \label{d19}%
\end{equation}
The density profile toward the drop center is proportional to
$\delta_{0}^{2}$ in contrast with the density near the contact line where it
is proportional to $\sqrt{\delta_{0}}$. This is consistent with our
previous findings (Eq. (\ref{d28})). For small $\delta_{0}$ majority of the
deposit thus occurs near the contact line.

We also note that toward the center of the drop ($l\rightarrow0$) it is
essentially in the very final stage of the whole drying process ($t\rightarrow
(1-\delta_{0}^{2})$). Practically some model assumptions may not hold at this
stage. For example, the drop height profile may no long decreases linearly in
time \cite{Deegan3}. However, the scaling property of $\Sigma\left(  r\right)
$ with respect to $\delta_{0}^{2}$ (Eq. (\ref{d19})) may still be valid.

\section{Deposit Properties with Diffusion-controlled Evaporation}

Evaporation profiles determine the hydrodynamics, such as the flow velocity
field, and thus affect the deposit properties.
Different evaporation profiles are achievable in experiments, and in some
cases the resulting deposit patterns change dramatically with evaporation
conditions \cite{Fisher1}. The uniform evaporation profile we have considered
is nonsingular while the diffusion-controlled evaporation profile that has
been generally assumed and studied in the literature diverges at the contact
line. We shall apply the pipe models to the deposition process with the
diffusion-controlled evaporation in this Section to assess the effects of
evaporation condition on general deposition properties.

\subsection{Hydrodynamics and Concentration Profile}

We derive the hydrodynamics and solute transport dynamics of an evaporating
thin drop with the same truncation concentration $C_{\max}$ but with the
diffusion-controlled evaporation profile.

Assuming vapor density above the liquid-vapor interface obeys the diffusion
equation and taking into account the quasi-equilibrium shape of the thin drop,
the evaporation flux defined as amount of mass per unit projected area per
unit time is%
\begin{equation}
J(r)\propto(1-r^{2})^{-1/2}, \label{B1}%
\end{equation}
where we still use the dimensionless representation: $r\rightarrow r/R$ and
$t\rightarrow t/t_{f}$ with $0\leq r\leq1$ and $0\leq t\leq1$.

The height profile takes the same form as with the uniform evaporation%
\begin{equation}
h(r,t)=\frac{\theta_{i}}{2}(1-r^{2})(1-t), \label{B3}%
\end{equation}
though physically in the dimensional domain the implicit total drying time
$t_{f}$ is defined with respect to the diffusion-controlled evaporation flux
as%
\begin{equation}
t_{f}=\frac{\rho\int_{0}^{R}2\pi h(r,0)rdr}{\int_{0}^{R}2\pi J(r)rdr}.
\label{B4}%
\end{equation}

Substituting Eqs. (\ref{B1}) and (\ref{B3}) into the condition of local
solvent mass conservation (Eq. (\ref{T7})), we solve for the velocity
field%
\begin{equation}
v(r,t)=\frac{1}{4}(1-t)^{-1}\frac{1}{r}\left[  \left(  1-r^{2}\right)
^{-1/2}-(1-r^{2})\right]  . \label{B2}%
\end{equation}
The height profile (Eq. (\ref{B3})) and the velocity field (Eq. (\ref{B2}))
thus completely determine the hydrodynamics.

To derive the solute concentration profile, we use the same method as with
the uniform evaporation (Eqs. (\ref{S3}) and (\ref{S7})). The
previously defined initial position function $r_{0}\left(  r,t\right)  $ (Eq. (\ref{S2})) and the
concentration $C(r,t)$ in the transport regime have exact though complicated
forms with the diffusion-controlled evaporation:%

\begin{equation}
\left(  1-r_{0}^{2}\right)  ^{3/2}=1-\left(  1-\left(  1-r^{2}\right)
^{3/2}\right)  \left(  1-t\right)  ^{3/4} \label{B8}%
\end{equation}
and
\begin{align}
C(r,t) & =C_{0}(1-t)^{-1/4}(1-r^{2})^{-1/2}\nonumber\\
& \times\left\{  1-\left[  1-\left(
1-r^{2}\right)  ^{3/2}\right]  (1-t)^{3/4}\right\}  ^{1/3}.\label{B5}%
\end{align}

The pipe models can be applied to derive the equation of motion of the moving
shock front%
\begin{equation}
\frac{dl}{dt}=-v\frac{\delta}{1-\delta},\label{B6}%
\end{equation}
where $v(r,t)$ is given by Eq. (\ref{B2}), and $\delta=C/C_{\max}$ with $C(r,t)$
given by Eq. (\ref{B5}).

At the contact line $dl/dt\rightarrow-v(1,0)\sqrt{\delta_{0}}/(1-\sqrt{\delta_{0}})$. It is finite
for the uniform evaporation, but divergent in the diffusion-controlled case (Eq. (\ref{B2})). 
We shall show later that this divergence does not affect the
deposit properties near the contact line.

As in the case of the uniform evaporation we can also introduce the
virtual front $\tilde{l}(t)$ by imposing the condition $C(\tilde{l}(t),t)=C_{\max}$, where
$C(r,t)$ is given by Eq. (\ref{B5}). The terminal time when the
virtual front arrives at the center of the drop is given by the
condition $C(0,t)=C_{\max}$, and we find

\begin{equation}
t=1-\delta_{0}^{4}.\label{B18}%
\end{equation}

Thus the whole transport process stops at $(1-\delta_{0}^{4})$ of the total
drying time for the diffusion-controlled evaporation as opposed to
$(1-\delta_{0}^{2})$ for the uniform evaporation.    

\subsection{Deposit Density Profile }

\begin{figure}
[ptb]
\begin{center}
\includegraphics[
width=0.88\columnwidth
]%
{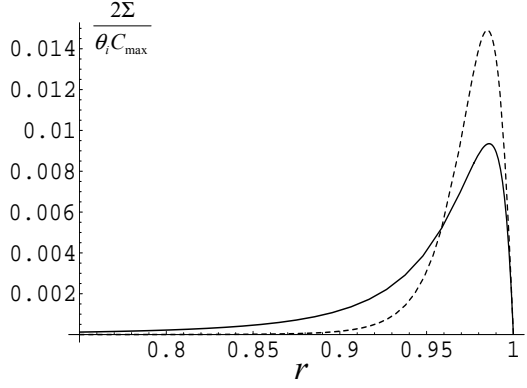}%
\caption[Deposition profile with diffusion-controlled evaporation]{Deposition profiles with $\delta_{0}=0.002$. 
The solid line corresponds to the case of the uniform evaporation. The dotted line corresponding 
to the case of the diffusion-controlled evaporation.}
\label{two_profiles}%
\end{center}
\end{figure}

\begin{figure}
[ptb]
\begin{center}
\includegraphics[
width=0.88\columnwidth
]%
{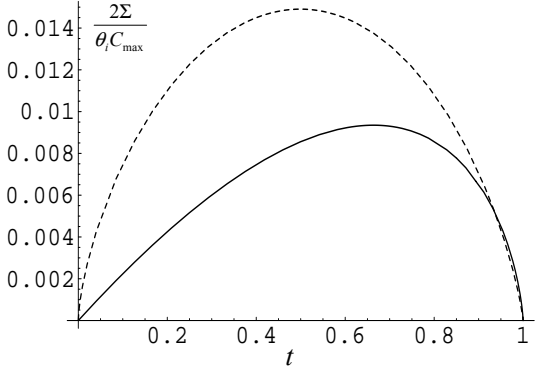}
\caption[Deposition growth dynamics with diffusion-controlled evaporation]{Deposition growth
dynamics with $\delta_{0}=0.002$. The solid line corresponding to the case of the uniform evaporation. The dotted
line corresponding to the case of the diffusion-controlled evaporation.}
\label{two_dynamics}
\end{center}
\end{figure}

For the case of the diffusion-controlled evaporation, the areal density
profile $\Sigma(r)$ and the growth dynamics $\Sigma(t)$ are completely solved
by Eqs. (\ref{d1}), (\ref{d20}), and (\ref{B6}). We show the numerical results
for the density profile and the growth dynamics in Fig. \ref{two_profiles} and
Fig. \ref{two_dynamics}. The corresponding results in the case of the uniform
evaporation are also shown for comparison.

The diffusion-controlled evaporation thus yields more pronounced deposit
density near the contact line as the singular flow field (Eq. (\ref{B2}))
carries more solute particles toward the contact line at early time. For the
areal density profile, two asymptotic regimes can again be identified as
detailed later, with majority of the deposit occurs near the contact line.
For the deposit growth dynamics, however, significant amount of deposit
occurs throughout the whole drying process. Thus the two different evaporation
conditions lead to density profiles and growth dynamics with qualitatively
similar properties.

We now study the asymptotic regimes quantitatively. Toward the contact line,
we consider the limit $r\rightarrow1$ and $\delta\rightarrow\delta_{0}\ll1$.
The equation of motion for the shock front  (Eq. (\ref{B6})) reduces to

\begin{equation}
\frac{dl}{dt}=-\frac{\delta_{0}}{8(1-t)}\frac{1}{1-l}\left(  1-t\right)
^{-1/4}\left(  1-\left(  1-t\right)  ^{3/4}\right)  ^{1/3}.\label{B14}%
\end{equation}
With the condition $l(0)=1$, Eq. (\ref{B14}) admits a solution of the form:%
\begin{equation}
\left(  1-l\right)  ^{2}=\frac{\delta_{0}}{3}\int_{(1-t)^{3/4}}^{1}%
(1-x)^{1/3}x^{-4/3}dx.\label{B9}%
\end{equation}

Using the scaled density profile $\tilde{\Sigma}$ (Eq. (\ref{d21})) and the
scaled contact line distance $\tilde{d}$ (Eq. (\ref{d22})), we can rewrite Eq.
(\ref{B9}) as
\begin{equation}
\tilde{d}^{2}=\frac{1}{3}\int_{\left(  \tilde{\Sigma}/\tilde{d}\right)
^{3/4}}^{1}(1-x)^{1/3}x^{-4/3}dx,\label{B10}%
\end{equation}
and%
\begin{equation}
\left(  \frac{\tilde{\Sigma}}{1-t}\right)  ^{2}=\frac{1}{3}\int_{(1-t)^{3/4}%
}^{1}(1-x)^{1/3}x^{-4/3}dx.\label{B11}%
\end{equation}
Eqs. (\ref{B10}) and (\ref{B11}) thus give the universal density profile and
growth dynamics in the limit $\delta_{0}\rightarrow0$.

Subjecting the growth dynamics Eq. (\ref{B11}) to the condition of vanishing
derivative with respect to time, we find that $t_{m}$, when the maximal areal density
occurs, satisfies the equation: 
\begin{align}
& \frac{2(1-t_{m})}{3}\int_{(1-t_{m})^{3/4}}^{1}(1-x)^{1/3}x^{-4/3}dx\nonumber\\
& -\frac{1}{
4}(1-(1-t_{m})^{3/4})^{1/3}(1-t_{m})^{3/4}=0. \label{B99}
\end{align}
Eq. (\ref{B99}) can be solved numerically, and thus
\begin{equation}
t_{m}=0.50...\label{B15}%
\end{equation}

It is worth noting that $t_{m}$ is not exactly $0.5$, and Eq. (\ref{B99}) is not exactly symmetric
about $t_{m}$ though the reverse might seem true from Fig. \ref{two_dynamics}.

Substituting $t_{m}$ into the shock front solution Eq. (\ref{B9}) and the
areal density profile Eq. (\ref{d1}), we find the contact line distance for
the maximal density
\begin{equation}
1-r_{m}=0.33...\sqrt{\delta_{0}},\label{B16}%
\end{equation}
and the maximal areal density
\begin{equation}
\Sigma_{m}=0.17...\sqrt{\delta_{0}}\theta_{i}C_{\max}\text{.}\label{B17}
\end{equation}

We also look for the power law governing the tail decay of the scaled universal
profile $\tilde{\Sigma}$. 
In the limit $\tilde{d}\rightarrow\infty$, the right side of Eq. (\ref{B10}) must diverge and the
integrand must be asymptotically dominated by $x^{-4/3}$ in the lower limit $(\tilde{\Sigma}/\tilde{d})^{3/4}\rightarrow0$.
Thus Eq. (\ref{B10}) reduces to the following power
law:\footnote{Alternatively, we can derive the power law exponent explicitly by using the formula $\int_{p}^{1}(1-x)^{1/3}x^{-4/3}dx=-3\left(
2(1-p)^{1/3}+pF_{1}^{2}[2/3,2/3,5/3,p]\right)  /(2p^{1/3})$. $F_{1}%
^{2}[2/3,2/3,5/3,p]$ is a hypergeometric function with $F_{1}^{2}%
[2/3,2/3,5/3,0]=1$ and $F_{1}^{2}[2/3,2/3,5/3,1]=\Gamma\left(  1/3\right)
\Gamma(3/5)$.}
\begin{equation}
\tilde{\Sigma}\sim\tilde{d}^{-7},\label{B12}%
\end{equation}
and in terms of the original variables $\Sigma$ and $r$
\begin{equation}
\Sigma\sim\theta_{i}C_{\max}\frac{\delta_{0}^{4}}{\left(  1-r\right)  ^{7}%
}.\label{B13}%
\end{equation}

In comparison with the uniform evaporation profile, a more pronounced maximal
areal density occurs earlier at about $1/2$ of the total drying time as
opposed to $2/3$ of the total drying time, but at about the same distance from
the contact line. The areal density profile near the contact line still obeys
the $\sqrt{\delta_{0}}$-scaling law. Away from the contact line, however, the
density profile decays much faster and is proportional to $\delta_{0}^{4}$ as
opposed to $\delta_{0}^{2}$ for the uniform evaporation.\footnote{It is
consistent with the previous results for the terminal time (Eq. (\ref{B18}))
when the whole transport process stops. This can be shown by considering the
concentration exactly at the center of the drop.} 

We are for the first time able to derive the power-law decay (Eqs. (\ref{d28}) and (\ref{B13})) 
of the deposition profile toward the center of the drop. Although we have based the derivation on specific
deposition mechanism and simple conditions, these asymptotic results reflect the behavior of the
shock front at a late time, and thus may reveal general properties of the evaporative deposition phenomena. We shall
come back to this point later in Discussion.   

\section{Comparison with Popov's Model}

As mentioned in Introduction, Popov \cite{Yuri3} applied the same
truncation criterion by Dupont to a similar problem with different conditions
and assumptions. Unlike our analysis so far, Popov studied the solute transport
process with diffusion-controlled evaporation
and analyzed the final deposit pattern in terms of the thickness profile.
These differences are not essential, however, as we shall discuss below.
Most importantly Popov assumed that the deposition region
interferes with the transport region and the hydrodynamics of the evaporating
drop. In particular, the formed deposit with finite thickness alters the boundary conditions
under which the drop shape is determined. In Popov's model the drop shape is thus truncated by the deposit: the drop shape
in the deposition region is identical to the thickness profile of the formed deposit while in the transport region it is
described by a spherical cap elevated by the amount of deposit thickness at the boundary.
As a result, the velocity field, determined by conservation of the solvent mass, also depends on the deposit
thickness, and the deposition process thus couples with the transport process.          
This assumption, though more realistic, makes the underlying mathematical structure more
complicated, and a complete solution is not available. Popov showed the asymptotic deposition properties 
near the contact line under these assumptions. We shall compare our findings with these results to 
shed some light on the model robustness.

\subsection{Thickness Profile}

\begin{figure}
[ptb]
\begin{center}
\includegraphics[
width=1.0\columnwidth
]%
{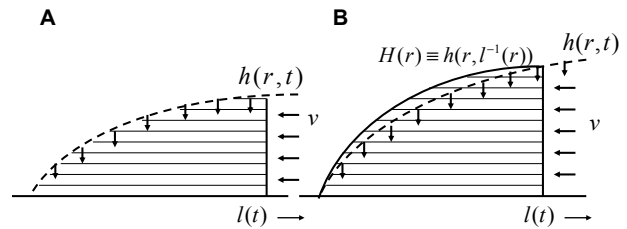}%
\caption [Areal density profile versus thickness profile]{Sketch of the two descriptions. The shaded
area is the deposition region. The dotted line outlines the height profile of
the evaporating drop that decreases in time. The solid line shows the thickness profile of the formed deposit.
A: description in terms of the areal
density profile where solute particles move vertically in the deposition region and the height profile of the region follows that of the drop; B: description in terms of the thickness profile
where the thickness of the deposit is described by $H(r)$.}
\label{thickness profiles}%
\end{center}
\end{figure}

We have described the final deposit pattern in terms of the areal density
profile. To implement this description, we assume that in the deposition
region, where horizontal transport is stopped, solute particles can still move
vertically so that the deposit thickness decreases along with the drop
height profile and eventually reaches zero. At the same time, the volume concentration diverges and is
replaced by the areal density profile.  
This zero-thickness assumption is not essential as long as the
deposition region does not interfere with the geometry and hence the
hydrodynamics of the evaporating drop. Alternatively we can interpret the
final deposit in terms of the thickness profile. In this description
solute particles cannot move vertically either once the volume concentration
reaches $C_{\max}$. Thus the thickness profile of the deposition region
does not change with time once formed and the volume concentration is uniformly $C_{\max}$ throughout the deposit. 
The evaporating drop, on the other hand, still maintains its spherical-cap equilibrium shape with height profile
decreasing in time, and the evaporation flux and the flow velocity field are
not affected by the deposit. Essentially these two descriptions correspond to two different deposit dynamics in the
deposit region during the drying process: in the areal density description, the deposit height profile is identical to the drop
height profile $h(r,t)$ and vanishes at the total drying time; in the thickness description, the density profile is characterized by the function $h(r,l^{-1}(r))$ and
is independent of time. We sketch these two descriptions in Fig. \ref{thickness profiles}.

The two mathematical formulations for the deposit profile are essentially the
same as the solute mass across the thickness at each position in the deposition
region is constant without horizontal transportation. The areal density
profile is given by $\Sigma\left(  r\right)  =C_{\max}h(r,l^{-1}\left(
r\right)  )$ (Eq. (\ref{d1})), and thus
\begin{equation}
H\left(  r\right)  \equiv\frac{\Sigma\left(  r\right)  }{C_{\max}}=h(
r,l^{-1}(r))  \label{D1}%
\end{equation}
is the corresponding thickness profile. Since $H(r)$ is proportional to
$\Sigma(r)$ with the constant coefficient $C_{\max}$, our previous findings
for $\Sigma\left(  r\right)  $ are readily applicable to $H\left(  r\right)$ with almost no modifications.

The mathematical equivalence is based on the same underlying physical assumption that the deposition process does not
interfere with the transport process and should be distinguished from Popov's model.  
This assumption may seem even farther from reality in the thickness description as finite accumulation of the deposit almost
certainly alters the velocity field and hence the transport process.
However, as we shall show later deposition properties are actually independent of the 
specific physical assumptions in some asymptotic regimes.  

\subsection{Asymptotic Deposit Properties}

We first consider, in the limit $t\rightarrow0$, $r\rightarrow1$, the slope of
the deposit thickness profile $dH/dr$. With $\Sigma$ given by Eq.
(\ref{d1}),%
\begin{align}
\frac{d\Sigma}{dr}  &  =-r\theta_{i}C_{\max}\left(  1-l^{-1}(r)\right)
-\frac{1}{2}\theta_{i}C_{\max}(1-r^{2})\frac{dl^{-1}(r)}{dr}\nonumber\\
&  =-\theta_{i}C_{\max}-\frac{1}{2}\theta_{i}C_{\max}(1-r^{2})\left(
\frac{dl}{dt}\right)  ^{-1}. \label{D2}%
\end{align}
In the limit the second term on the right side of Eq. (\ref{D2}) is
proportional to $h(1,0)/v(1,0)$ at the contact line and vanishes. Thus near
the contact line in the early drying stage $d\Sigma/dr\rightarrow-$
$\theta_{i}C_{\max}$ and thus $dH/dr\rightarrow-\theta_{i}$, which is
independent of the evaporation profile (compared with Eq. (\ref{F25})).

We next consider the growth dynamics of the deposit thickness
$H(t)\equiv\Sigma\left(  t\right)  /C_{\max}$ at the early drying stage. Since
$H\left(  r\right)  $ as shown above is proportional to $\left(  1-r\right)  $
in this limit, thus
\begin{equation}
H(t)=H\left(  r\right)  \rightarrow\theta_{i}\left(  1-l\left(  t\right)
\right)  . \label{D3}%
\end{equation}
For the uniform evaporation, it is given by Eq. (\ref{F5}). For the
diffusion-controlled evaporation we note that near the contact line
$dl/dt\rightarrow-\sqrt{\delta_{0}}v(l,t)$, and thus in this limit
\begin{equation}
1-l(t)\approx\sqrt{\delta_{0}}\left(  1-r_{0}(1,t)\right)  , \label{D4}%
\end{equation}
where $r_{0}\left( r,t\right)$ is the initial position function (Eq. (\ref{B8})).
\ With $r=1$, $t\rightarrow0$, and $r_{0}\rightarrow1$, the leading order term
of $H(r)$ (Eq. (\ref{B8})) is 
\begin{equation}
H\left(  t\right)  \approx\sqrt{\delta_{0}}\theta_{i}\frac{\left(  3t\right)
^{2/3}}{2^{7/3}}, \label{D6}
\end{equation}
using $\left(  1-r_{0}^{2}\right)  \approx2\left(  1-r_{0}\right)  $
and $1-r_{0}\approx (1/2)(1- (1-t)^{3/4})^{2/3}\approx (3t)^{2/3}/{2^{7/3}}$.

The asymptotic form Eq. (\ref{D6}) is exactly the same as the result derived
in Popov's case \cite{Yuri3}. This sameness is not a surprise. In solving the
system of coupled equations, Popov argued that the solutions can be expanded in
terms of $\sqrt{\delta_{0}}$. In the lowest order the equations
decouple,\footnote{The solution Eq. (\ref{D6}) corresponds to the zeroth-order
in $\sqrt{\delta_{0}}$ in the asymptotic expansion \cite{Yuri3}. \ The
explicit dependence on $\sqrt{\delta_{0}}$ in Eq. (\ref{D6}), however, appears via a
different mathematical route. In Dupont model, it is due to the boundary condition Eq.
(\ref{P17}) at the contact line.} and it physically corresponds to the
condition that the deposition process and the hydrodynamics are separated.
Thus our result Eq. (\ref{D6}) (as well as the assumption of decoupling) is
the lowest order approximation to Popov's findings near the contact line at the
early drying stage.

The later drying stage when $t$ is large, however, was not well defined in
Popov's results. Popov suggested that, in the limit $\delta_{0}\rightarrow0$,
the thickness profile $H(r)$ monotonically increases with $t$ and eventually
approaches a limit $H_{m}$ that is proportional to $\sqrt{\delta_{0}}$ toward
the end of the drying process ($1-t\ll1$). The thickness profile thus ends
abruptly with a \textquotedblleft vertical wall\textquotedblright\ at its
inner side. Although our finding of the maximal thickness $H_{m}\equiv
\Sigma_{m}/C_{\max}$ has the same $\sqrt{\delta_{0}}$-dependence (Eqs.
(\ref{d15}) and (\ref{B17})), the universal properties of the profile show that $H_{m}$ is
formed at about $2/3$ (uniform evaporation) or $1/2$ (diffusion-controlled evaporation) 
of the total drying time and the thickness decreases
continuously afterward following a power-law decay (Eqs. (\ref{d28}) and (\ref{B13})). Furthermore, our results clearly show two different
asymptotic regimes and the thickness profile is instead proportional to
$\delta_{0}^{2}$ (uniform evaporation) or $\delta_{0}^{4}$ (diffusion-controlled evaporation) 
in the late drying stage toward to center of the drop.
Neither the $\sqrt{\delta_{0}}$-universality of the thickness profile near the
contact line nor the power-law decay toward the center of the drop was
apparent in Popov's findings.

The coupled system Popov studied is thus fully reducible to the simpler
conditions we have assumed in the small $\delta_{0}$ limit. This coupling
between the deposition and the hydrodynamics is further resolved near the
contact line in the limit $r\rightarrow1$ (this is consistent with the
$\sqrt{\delta_{0}}$-scaling property of $(1-r)$). It is thus fairly reasonable
to argue that the evaporative deposition phenomenon is robust against specific
assumptions and conditions in this asymptotic regime. Additionally as we have
shown that the temporal properties of the process are actually $\delta_{0}%
$-independent, by assuming the decoupling we are able to specify a universal
time scale when the maximal deposit thickness forms and to extend
the truncated deposition model to a separate new asymptotic regime toward the
center of the drop.

Furthermore, it is worth pointing out that even Popov's treatment in \cite{Yuri3} only partly addresses the coupling
between the formed deposit and hydrodynamics, and a few important dynamical aspects are still missing. First, the
deposit-dependent drop shape alters the boundary conditions under which the evaporation flux is determined and hence further
changes the velocity field. Second, the velocity field in the deposition region should be certainly affected by the deposit.
If the deposit takes up a nonzero fraction of the volume, a given evaporation flux will take the same amount of 
available solvent per unit time but with a larger effective volume in the deposition region, and a larger velocity is required 
to replenish this solvent. However, our findings suggest that by controlling the initial concentration ration $\delta_{0}$ 
all these couplings might be effectively suppressed near the
contact line at the early drying stage and may not alter the deposition properties. Also, further
considerations of these couplings are important for experimentally testing the model. We shall
discuss this issue later.
    
\section{Comments on Depinning and Multiple-Ring Patterns}

In the evaporative deposition model, we have identified one key
deposition mechanism: horizontal transport of solute particles truncated by
maximal allowed volume concentration. This truncated dynamics is robust in the sense 
that it admits a solution with a moving shock front that has a definite functional form 
independent of the dynamical characteristics of the problem. 
The shock front velocity only depends on local values of the flow velocity, 
the pipe geometry, and the volume concentration in the transport region. 
Furthermore, the truncated transport
mechanism leads to a characteristic deposit profile near the contact line that is 
independent of evaporation conditions and dynamical couplings in the early drying stage.

Although practically evaporative deposition phenomena are rich and there must
be many complex underlying mechanisms, we now discuss how the truncated
transport dynamics, as a single deposition mechanism, might yield general
deposit profiles under some typical and controllable conditions.

\subsection{Depinning Configuration}

For an actual evaporative deposition process \cite{Deegan3}, in the late
drying stage the contact line of a evaporating drop can depin and retreat
toward the drop center before it is pinned again, and the deposition process
is restored at the newly-formed contact line. The slip-stick process may occur
many times before the drop tries up and thus yield a multiple-ring pattern
\cite{Stone2, Jia1, Xu1, Hong1}.

There is still no consensus on a general depinning mechanism, which may not exist after all. 
A conventional explanation \cite{Deegan3, Yuri3, Nonomura1} calls for the Young-Dupre equation at the contact line
\cite{Witten1}: $\gamma_{lv}\cos\theta+\gamma_{sl}-\gamma_{vs}=0$, where
$\theta$ is the contact angle, and $\gamma_{lv}$, $\gamma_{sl}$, and
$\gamma_{vs}$ are the surface tensions at liquid-vapor, solid-liquid, and
vapor-solid interfaces. The horizontal component of the surface tension along
the liquid-vapor interface at the contact line $\gamma_{lv}\cos\theta$ acts as
the depinning force that increases as the drop progressively flattens toward
the late drying stage. $\gamma_{sl}$ also contributes to the depinning force,
but it primarily depends on the local deposition and is less affected by the geometry.

Admittedly this explanation can be problematic in that it predicts an extremely narrow range
of deposit thickness and drop configurations when the maximal horizontal depinning force is
achieved and the contact line depins. In reality, however, the depinning occurs over 
a wide range of configurations. To account for this smooth change, the conventional explanation thus must be complemented by
other mechanisms that are still lacking. We shall in the following consider within the framework of the conventional
theory and cautiously keep its limitations in mind.  

The depinning phenomenon thus clearly involves the coupling between the
deposition process and the hydrodynamics of the evaporating drop: the drop
geometry and therefore the contact angle $\theta$ must be determined from the
boundary conditions defined by the formed deposit pattern to account for
the change of the depinning force (Fig. 1.2A). To continue our discussion of
depinning within the model established in the previous Sections, we must adopt
the view that our results based on the assumption of decoupling can be applied to the
coupling system in the lowest order of $\delta_{0}$, and we also must
interpret the deposit pattern in terms of the thickness profile $H$.

Deegan showed experimentally \cite{Deegan3} that the depinning time $t_{d}$ is
about $0.4$ to $0.8$ (in the unit that the total drying time $t_{f}=1$) and it
depends on the initial volume concentration $\delta_{0}$ (from $0.001 $ to
$0.01$). We have found that in the small $\delta_{0}$ limit, the maximal areal
density and hence the maximal deposit thickness occurs at time
$t_{m}\approx0.67$ (Eq. (\ref{d13})) for the uniform evaporation and $0.50$ (Eq. (\ref{B15})) 
for the diffusion-controlled evaporation. 
Our $t_{m}$ thus falls in the range of
the depinning time found in experiments. Geometrically the maximal deposit thickness
occurs when the contact angle $\theta$ at the rim of the deposit becomes
zero and the depinning force $\gamma_{lv}\cos\theta$ achieves its maximal value.

To derive $t_{m}$ formally, we note that $H(t)=\Sigma\left(  t\right)
/C_{\max}=(\theta_{i}/2)h(l(t),t)$ (Eq. (\ref{d20})), and thus the condition
$dH/dt=0$ demands%
\begin{equation}
\frac{\partial h}{\partial t}+\frac{\partial h}{\partial r}\frac{dl}{dt}=0.
\label{D7}%
\end{equation}
Eq. (\ref{D7}) does not depend on model assumptions as $h(r,t)$ can be the
drop height profile under any conditions. In particular, $h(r,t)$ can be the
solution to the drop geometry coupled with the deposition process. $dl/dt $,
on the other hand, must be given by the pipe model results Eq. (\ref{P7}).
Eq. (\ref{D7}) gives the general condition and scheme to solve for $t_{m}$
when the maximal thickness occur within the truncated transport model.

However, $t_{m}$ is only an approximation for the depinning time $t_{d}$. The
depinning may occur once the contact angle is small enough but not necessarily
zero, and as a result $t_{d}\leq t_{m}$.\footnote{As noted by Popov
\cite{Yuri3}, the contact angle $\theta$ may become negative when the drop is
actually overhanging against the deposit rim. In this case $t_{d}>t_{f}$,
and the depinning process must be due to different mechanisms.} More
importantly, our finding of $t_{m}$, as opposed to Deegan's results, does not
depend on the initial concentration (Fig. \ref{scaling law_t}) when
$\delta_{0}$ is small. This dependence may be included in the higher order
solution in $\delta_{0}$ when the coupling between deposition and geometry is
considered. The dependence can also reflect the dependence of $\gamma_{sl}$ on
$\delta_{0}$. Nevertheless our finding of $t_{m}$ within the truncated
transport model is consistent with empirical evidence and may be instructive
for further studies.

Also, for the complicated depinning process the above explanation based on
Deegan's experimental results is not universal. Other mechanisms, such as the
tension within the formed deposit \cite{Gennes2}, the kinetic processes
controlled by the viscous stress in the drop, or the diffusion relaxation in air, 
may affect the depinning under certain conditions and thus lead to different 
interpretation of the depinning time.

\subsection{Formation of Multiple-Ring Patterns}

The depinning process can lead to multiple-ring deposit patterns: the
retreating contact line is rested at a nearer position toward the drop center
and the deposition process continues at this newly formed contact line. In
Refs. \cite{Adachi1} and \cite{Nonomura1}, a slip-stick theory was
established, where the Young-Dupre condition was incorporated into the
equation of motion for the contact line. As a result, the contact line was
found to follow a sinusoidal motion, and this led to a very dense
multiple-ring pattern. The authors in Refs. \cite{Stone2} and \cite{Xu1}
considered controllable multiple-ring formation in nonconventional geometries. Some of these
empirical evidence may need explanations that are different from Deegan's
framework \cite{Deegan3}.

Following our discussion on the depinning process, 
we consider an ideal model
for the multiple-ring formation. For the diffusion-controlled evaporation, we identify the depinning time $t_{d}
=t_{m}\approx1/2$ ($2/3$ for the uniform evaporation), \textit{i.e.}, the contact line is released when the
contact angle becomes zero. The width of the deposit ring is identified as
$\left(  1-l_{m}\right)  $ (Eq. (\ref{d14})). The thickness is represented by
the maximal thickness $H_{m}=\Sigma_{m}/C_{\max}$ (Eq. (\ref{d15})).

The newly pinned position $R^{\prime}$ can be determined by considering the
change of the solvent volume. Initially for small contact angle $\theta_{i}$,
the total volume of the drop is
\begin{equation}
V\approx\frac{\pi}{4}R^{3}\theta_{i}\text{.} \label{D8}%
\end{equation}
We assume the contact line moves fast enough to the new position so that the
time it takes can be ignored compare to the time when it is pinned. Since the
evaporation flux is stationary, the volume of the drop when the contact line
is pinned again is thus $V(1-t_{d})\approx V/2$ ($V/3$ for the uniform evaporation). For slow evaporation, the newly
pinned drop must still maintain a spherical cap geometry. The contact angle at
the new contact line must still be approximately $\theta_{i}$ since it is
determined by the Young-Dupre condition. Thus we find%
\begin{equation}
R^{\prime}\approx\left(\frac{1}{2}\right)^{1/3}R\approx0.8R\text{,} \label{D9}%
\end{equation}
and $R^{\prime}\approx0.7R$ for the uniform evaporation.

We next consider the width and the thickness of the new deposit ring. The
deposition process via truncated transport is restored at the newly formed
contact line, and the deposit growth there must follow what we have found
at the initial contact line. Since $R(1-r_{m})\propto R\sqrt{\delta_{0}}$ (Eq.
(\ref{d14})) and $H_{m}\propto R\sqrt{\delta_{0}}$ (Eq. (\ref{d15})) (we now
have added $R$ to the dimensionless expressions), for the new ring we replace
$R$ with $R^{\prime}$ and replace $\delta_{0}$ with concentration ratio at
the new contact line: $\delta_{0}^{\prime}\approx C(R^{\prime},t_{d})/C_{\max}\approx2.2\delta_{0}$ in the case of the diffusion-controlled evaporation 
profile (Eq. (\ref{S7})) ($2.4\delta_{0}$ for the uniform evaporation). Thus the width $R^{\prime}(1-r_{m}^{\prime})$
and the thickness $H_{m}^{\prime}$ of the new ring satisfies:%
\begin{equation}
R^{\prime}(1-r_{m}^{\prime})\approx1.1R(1-r_{m}), \label{D11}%
\end{equation}%
\begin{equation}
H_{m}^{\prime}\approx1.1H_{m}\text{.} \label{D12}%
\end{equation}
For uniform evaporation, they are $1.2R(1-r_{m})$ and $1.2H_{m}$ respectively.
The thickness and height of the secondary ring are thus comparable to those of
the initial ring.

There can be further depinning events in this ideal model. Following the above
reasoning, for the $n$-th order ring (the initial ring is the zeroth order
with the radius $R$) toward the drop center, the time it starts to form (the
depinning time of the $\left(  n-1\right)  $-th ring), the radius, and the
concentration ratio are%
\begin{equation}
t_{d}^{\left(  n-1\right)  }\approx\frac{1}{2}\sum_{k=0}^{n-1}\left(  \frac
{1}{2}\right)  ^{k}=1-\frac{1}{2^{n}}\text{,} \label{D15}%
\end{equation}%
\begin{equation}
R^{(n)}\approx\left(  \frac{1}{2}\right) ^{n/3}R\text{,} \label{D13}%
\end{equation}
and
\begin{equation}
\delta_{0}^{(n)}\approx\frac{C(R^{(n)},t_{d}^{(n-1)})}{C_{\max}}%
=\frac{2^{7n/6}-1}{2^{2n/3}-1}\delta_{0}\text{.} \label{D14}%
\end{equation}
Thus the width and the thickness of the $n$-th ring are%
\begin{equation}
R^{(n)}(1-r_{m}^{(n)})\approx\left(  \frac{1}{2}\right)  ^{n/2}\left(
\frac{2^{7n/6}-1}{2^{2n/3}-1}\right)  ^{1/2}R(1-r_{m})\text{,} \label{D16}%
\end{equation}%
\begin{equation}
H_{m}^{(n)}\approx\left(  \frac{1}{2}\right)  ^{n/2}\left(  \frac{2^{7n/6}%
-1}{2^{2n/3}-1}\right)  ^{1/2}H_{m}\text{.} \label{D17}%
\end{equation}
The results for the uniform evaporation can be derived similarly.

\begin{figure}
[ptb]
\begin{center}
\includegraphics[
width=0.90\columnwidth]
{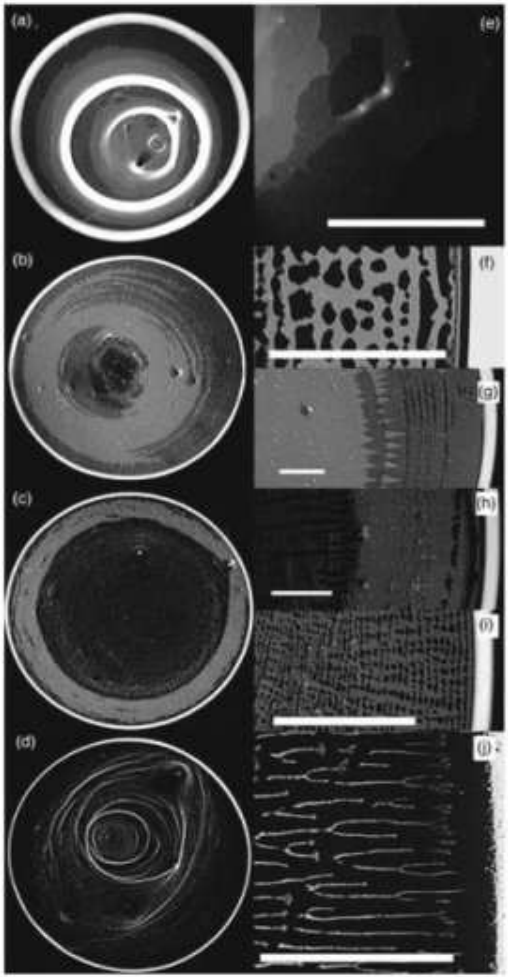}
\caption[Photographs of multiple-ring patterns in Deegan's experiments]{Photographs of the deposit patterns in Deegan's experiment
\cite{Deegan3}. The left column shows the entire deposit for initial volumn
concentration (a) 0.01, (b) 0.0025, (c) 0.0013, (d) 0.00063. The right column
shows the close-ups. (Courtesy Robert Deegan)}%
\label{deegan}%
\end{center}
\end{figure}

Empirically deposit patterns with major multiple rings were observed by
Deegan \cite{Deegan3}. In Fig. \ref{deegan}, for two initial concentrations
(a) 0.01 and (d) 0.00063 there are major higher-order rings up to the third
order with width and thickness comparable to the zeroth order ring at the
initial contact line. Our above analysis in the highly ideal case is at least
consistent with the observed pictures. For the other two initial
concentrations (cases (b) and (c)), however, major higher-order rings are not observed.

Between the major rings Deegan's results clearly show some fine complex
patterns. In our ideal case, these fine patterns are formed by the solute
particles in the region where the depinned contact line sweeps through between two
pinning events. 
Thus they must be formed within a much smaller time scale compared to the major
ring formation according to our assumptions.\footnote{We estimated in
\cite{Zheng2} that the typical velocity for a free standing dewetting film is about the
order of 1 m/s.} Formation of those fine
structures, as a result, may not be explained within the framework of the
truncated transport dynamics considered in this work. Instead, other
mechanisms, such as the dynamics and instabilities of the dewetting of
thin films \cite{Gennes3, Ohara1, Nakanisi1} where capillary forces are important and
long-range interactions between solute particles are relevant \cite{Brenner1,
Nierop1}, are needed to account for the observed patterns in those regimes.

\section{Discussion}

\subsection{Experimental Aspects of the Model}

Experimental tests involve two aspects of the model. First, Dupont model is based on some
assumptions and conditions that require specific experimental design and realization. 
Second, testability depends on the major predictions made by the model. We shall discuss
these in turn.

The decoupling between the deposition process and the transport process is the major
assumption. This ideal case can be approximated under certain experimental conditions.

One of the key parameters in Dupont model is the allowed maximal concentration
$C_{\max}$. $C_{\max}$ should depend on the size of the solute particles,
interactions between the solute, the solvent, and the
substrate. $C_{\max}$ could be also affected by other dynamic factors, such as
the flow velocity. Empirical data for $C_{\max}$ is limited, but naively
$C_{\max}$ can be deduced from the initial volume concentration by
measuring the volume of the major deposit rings and comparing it with the
initial volume of the entire drop.
 
The assumption of decoupling requires $C_{max}$ be small. Any solid particles must become
immobilized if their concentration is high enough. For simply shaped particles, the immobilization
occurs at volume fractions of order unity. However, the immobilizing volume fraction can be
much smaller than unity if the solute consists of tenuous objects like polymers, especially
those that form a physical gel \cite{Witten1}. Many organic solutes such as food may
well be of this sort, and seem most appropriate for the model.

Another aspect of the coupling is that the formed deposit may alter the drop geometry which
in turn changes the evaporation flux and its generated flow field. One way to bypass this
obstacle is to realize the uniform evaporation profile as sketched in \cite{Zheng1}. The
uniform evaporation, unlike the diffusion-controlled case, is maintained by the peripheral 
environment but not the drop itself. The coupling due to change of the evaporation condition is thus
effectively avoided.

Furthermore, some kinetic effects such as Marangoni flows and viscous stress may deviate from the Dupont model assumptions
and the fundamental lubrication approximation. Convective Marangoni flows can be avoided by
maintaining uniform experimental conditions such as the temperature and by reducing the thermal
effect due to evaporation. Viscous stress effect can
be also reduced by slowing the evaporation. 
Thus by making the environment nearly saturated with the
solvent vapor, these kinetic effects can be made arbitrarily small.

Our work allows one to compare the predictions of the Dupont model with experiment
quantitatively in several new ways. Our main prediction is about the normal density profile.
There are three aspects that might be tested: the leading edge, the time and position of
the maximal deposition, and the trailing edge.

The leading edge prediction states that the areal density should increase linearly with
distance from the contact line, and that the slope is proportional to the square root of
the initial concentration ratio. This prediction is identical to Popov's results and is therefore
not a particular test of Dupont model. The maximal deposition time, on the other hand, 
is a strong test and comes out very clearly from our findings.

The power-law decay of the areal density at the trailing edge is also a strong test. On the
qualitative level, this decay appears consistent with some experiments but not others. Often with
colloidal solutes, such as in some of Deegan's experiments, one sees a sharp inner edge with
no apparent decay at all. This is not a surprise. As noted above, Dupont model is not
completely applicable with compact solid solutes. Moreover, a continuum model such as Dupont's
should not work completely when the solute particles have a size comparable to the drop
thickness $h$.For non-colloidal solutes, there should be a decay that is consistent with our
predictions for $C_{max}$ small enough.

For quantitative prediction, one should measure the density or thickness of the deposit as
a function of distance as well as drying time. The function of distance can be
straightforwardly measured for the dried deposit. The function of drying time can be measured
by monitoring and recording the whole drying process. The shock front velocity might be
also measured this way.       

\subsection{Generalizations of the Model}

We have not considered any kinetic effects such as visous stress and diffusion in the work. These effects will modify 
the physical assumptions and mathematical formulations. For example, we implement a step function to model the deposition
condition with respect to the threshold concentration $C_{max}$. Horizontal diffusion and forces between various media may smooth
the transition to immobility. Mathematically this effect can be modelled by replacing the step function with a weakly
convergent series, and the truncated transport equation thus becomes nonsingular. This is an interesting mathematical problem.
Physically we want to point out that the smooth effect is not as necessary as it might seem. When $\delta_{0}$ is small, the concentration in the transport region rises
so fast toward the contact line that the transitional region with partial mobility would be very small.  

We already suggested that the power-law decay (Eqs. (\ref{d28}) and (\ref{B13})) of the trailing edge could be independent of
specific assumptions and derivation, and may actually reveal 
general properties of the evaporative deposition without depinning.
To be more specific, as already partly discussed in the previous work \cite{Zheng1} asymptotic properties, such
as the power-law decay, might be robust against a broad range of specific dynamic processes, but are instead dependent
on the singularities that characterize the dynamic conditions. In the case of the evaporative deposition, there are two
sources of singularities encountered: the vanishing of the drop height and the divergence of the diffusion-controlled
evaporation profile. Both occur at the contact line. The exponent of -7 in the power-law decay of the density profile might
be uniquely determined by these two singularities. This thought may prompt further studies of the robustness of the
power-law against the drop geometry and other singular conditions. For instance, an evaporating drop pinned in an angular
region studied in \cite{Zheng1} has a singular geometry profile at the contact line dependent on the contact line distance
from the vertex of the angle as well as the opening angle of the region. An exact relation between the decay exponent and
the characteristic singularities would give a decaying profile that continuously varies along the contact line and is
controllable by the opening angle. More detailed asymptotic analysis is needed toward this goal.

Although in this work the simple geometry (spherical cap) and the regular
evaporation profiles (uniform or diffusion-controlled) yield consistent and
rich deposition phenomena, controllable nonconventional geometry and
evaporation (and so is the flow velocity field) are important for creating
complex deposit patterns \cite{Cranick1, Xu1, Hong1}. Rigorous theoretical
studies on these patterns are challenging. For example as pointed out by Popov
\cite{Yuri3}, determining the evaporation profile with nonconventional
boundary geometry is one of the major obstacles for rigorously solving the
deposition problem. Furthermore, some drop configurations and velocity field,
such as the negative contact angle, might lead to the formation of
higher-order deposit rings even before depinning happens.

The pipe models we have established in this work capture the basic
characteristics of the truncated transport dynamics. It is robust in the sense
that it admits a moving shock front, of which the equation motion only depends
on local characteristic quantities in a definite functional form. Besides the
evaporative deposition problem, pipe models may be relevant to other
systems with similar properties, such as the jamming in granular flow
\cite{Nagel1, Nagel2}. Further studies are needed to adapt the pipe models to such
systems.

In addition, some aspects of the pipe models may invite further attention. For example, 
to understand the behaviors of the shock front at singularities of a pipe 
and to characterize and utilize such behaviors require our further efforts. We have only
considered in this work the singularity associated with the vanishing height
of the pipe (Eq. (\ref{P17})), and in the application to the evaporative deposition problem this
singularity corresponds to the contact line and amounts to the condition of
self-consistency (see Appendix for details). However, the formulation Eq. (\ref{P17}) is far
more general than the application. Unlike the case of evaporative thin drops where the height
profile and the flow velocity field are coupled due to the mass conservation of the solvent,  
In our derivation of the pipe models, the height profile and the flow 
velocity field are allowed to have independent dynamics. Thus the singularities can rise either dynamically
as those characteristic quantities evolve or statically in a controllable way. How to describe the shock front with
dynamical singularities is an interesting but challenging problem.  

\section{Conclusion}

We have studied in this work a highly idealized evaporative deposition model in which
different dynamics is fully separatable. In addition, the model is scalable in the sense that
the length scale, defined in terms of the initial drop radius, and the time scale, defined in terms of the total
drying time, are irrelevant in model characteristics. Yet this model has yielded rich and solvable
deposition properties that are consistent with empirical evidence.  

In retrospection, we have so far fairly successfully addressed the two goals
suggested in the Introduction Section.

The first goal is to break the problem down and to prioritize different
components. We divide the contact line deposition problem into two parts:
phenomenology and mechanism. The former includes various conditions such as
evaporation profiles, capillary-dominant equilibrium drop shape, and flow
velocity field. For the latter part, among all the complexities we have
identified and focused on the truncated transport dynamics as the unique
deposition mechanism. Our findings suggest that this unique mechanism is
effective and robust for a fairly wide range of phenomenologies to produce
consistent deposit patterns.

Our analysis has revealed a few asymptotic regimes with different
characteristic scales in space and time. These regimes are largely determined
by the phenomenologies but not the mechanism. The truncated transport dynamics
is applicable to some of the regimes, and asymptotic deposit properties and
scaling laws are identified. But some other important regimes may only be
addressable by other mechanisms that go beyond the scope of this work.

The second goal is to find the appropriate mathematical structure, which is
largely determined by the mechanism. We have found the \textquotedblleft pipe
models\textquotedblright\ and the characteristic moving shock front solution
useful structures to describe the truncated transport dynamics and the
deposition problem. It has a universal functional form and depends only on
local conditions specified by the phenomenologies. It is adaptable and generalizable.

In summary we thus have found: transport dynamics truncated by the allowed
maximal volume density, combined with the hydrodynamics under lubrication
approximation of a thin drop evaporating regularly with uniform or
diffusion-controlled evaporation flux, gives a simple, tractable, and yet rich
enough picture for the general contact line deposition phenomena. A few
asymptotic regimes both in drying time and radial distance are identified and
analyzed. We are able to predict in certain cases deposit
profile and growth properties that are consistent with empirical evidence and
previous findings. These properties are largely robust against physical
conditions and theoretical assumptions, and are instructive to further applications.

\begin{acknowledgments}

This work is done as part of the fulfillment for the author's doctoral degree under the
supervision of professor Thomas A. Witten at the University of Chicago. The author pays great
gratitude to professor Witten for his constant guidance, encouragement, and understanding.
The author also thanks Professor Todd F. Dupont for useful comments. This work was supported
in part by the MRSEC Program of the National Science Foundation under Award Number DMR-0213745. 

\end{acknowledgments}

\appendix

\section{Direct Derivation of the Shock Front Velocity}

\begin{figure}
[ptb]
\begin{center}
\includegraphics[
width=0.9\columnwidth
]{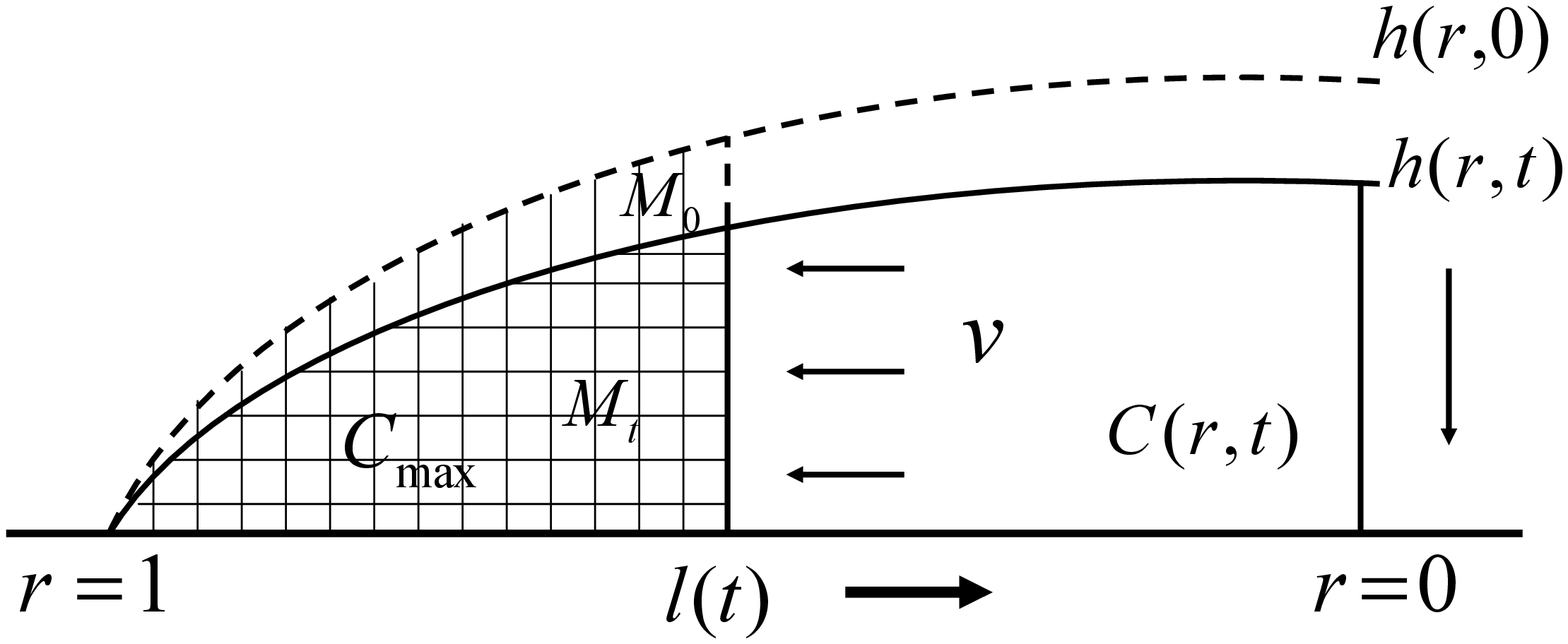}
\caption[Sketch of the definition of $M_{0}$ and $M_{t}$]{Sketch of the definition of $M_{0}$ and $M_{t}$. }%
\label{phase boundary}
\end{center}
\end{figure}

The shock front $l(t)$ moves from the contact line ($r=1$) toward the center
of the drop ($r=0$). Behind the boundary $1\leq r\leq l(t)$ is the
deposition region; in front the boundary, which is in the transport region,
solute particles are carried by the flow toward the boundary (Fig.
\ref{phase boundary}).%

We consider the change in total solute mass in the region $l(t)\leq r\leq1$.
Initially the amount of solute mass is%
\begin{equation}
M_{0}=\int_{l(t)}^{1}2\pi C_{0}h(r,0)rdr. \label{S8}%
\end{equation}
For a position $r$ with $l(t)\leq r\leq1$ when the moving front reaches $r$ at
time $l^{-1}(r)$, the volume concentration at $r$ becomes $C_{\max}$ and the
height at $r$ is $h(r,l^{-1}(r))$. The amount of solute mass near $r$ at time
$l^{-1}(r)$ is thus $2\pi rC_{\max}h(r,l^{-1}(r))dr$. Since there is no
horizontal movement of solute particles in the deposition region, this amount
of mass does not change in time. Therefore at time $t$ the total amount of
solute mass in the region $l(t)\leq r\leq1$ is%
\begin{equation}
M_{t}=\int_{l(t)}^{1}2\pi C_{\max}h(r,l^{-1}(r))rdr. \label{S9}%
\end{equation}
 
Conservation of solute mass demands that the change in total mass $M_{t}%
-M_{0}$ must be equal to the amount mass transported through $l(t)$ from time
zero to time $t$, \textit{i.e.},
\begin{equation}
\int_{l(t)}^{1}2\pi C_{\max}h(r,l^{-1}(r))rdr-\int_{l(t)}^{1}2\pi
C_{0}h(r,0)rdr=m(l(t),t), \label{S10}%
\end{equation}
$m(l(t),t)$ is defined by Eq. (\ref{S3}).
We differentiate against $t$ on both sides of Eq. (\ref{S10}),%
\begin{equation}
-2\pi lC_{\max}h(r,t)\frac{dl}{dt}+2\pi lC_{0}h(r,0)\frac{dl}{dt}%
=\frac{\partial m}{\partial t}+\frac{\partial m}{\partial l}\frac{dl}{dt},
\label{S11}%
\end{equation}
and solve for $dl/dt$:%
\begin{align}
\frac{dl}{dt}  &  =\frac{\partial m/\partial t}{2\pi l\left[  C_{0}%
h(r,0)-C_{\max}h(r,t)\right]  -\partial m/\partial l}\nonumber\\
&  =-\left[  \frac{l}{4(1-t)}\right]  \frac{1}{\frac{C_{\max}\left(
1-l^{2}\right)  }{C_{0}\left[  (1-t)^{-1/2}-l^{2}\right]  }-1}. \label{S12}%
\end{align}

Eq. (\ref{S12}) can also be written in the form:%
\begin{equation}
\frac{dl}{dt}=-v\left(  l,t\right)  \frac{\delta}{1-\delta}, \label{A10}%
\end{equation}
and we thus have re-covered Eq. (\ref{P16}).

The initial condition Eq. (\ref{P18}) can be proved via asymptotic expansion.
Near the contact line $t$ is small, and we can write $l(t)=1-\alpha
t+O(t^{2})$. Substituting this expression for $l(t)$ on both sides of Eq.
(\ref{P16}) and retaining only the lowest order of $t$ in the limit
$t\rightarrow0$, we find
\begin{equation}
\alpha=\frac{1}{4}\frac{1}{\frac{4\alpha}{\delta_{0}(1+4\alpha)}-1},
\label{A11}%
\end{equation}
We solve for $\alpha$:%
\begin{equation}
\alpha=-\left.  \frac{dl}{dt}\right\vert _{t=0}=\frac{1}{4}\frac{\sqrt
{\delta_{0}}}{1-\sqrt{\delta_{0}}}. \label{A12}%
\end{equation}
The initial condition Eq. (\ref{P18}) is thus proved.

Eqs. (\ref{S12}) and (\ref{A12}) completely define the equation of the motion
for the shock front.

\end{document}